\documentclass[prd,eqsecnum,twocolumn,amsfonts,amssymb]{revtex4}

\usepackage{graphicx}

\usepackage{bm}

\newcommand{\beq}{\begin{equation}}
\newcommand{\eeq}{\end{equation}}
\newcommand{\beqs}{\begin{eqnarray}}
\newcommand{\eeqs}{\end{eqnarray}}

\begin{document}

\title{Large-$N_c$ and Large-$N_F$ Limits of SU($N_c$) Gauge Theories with 
Fermions in Different Representations}

\author{Sudhakantha Girmohanta$^a$, Thomas A. Ryttov$^b$ and 
Robert Shrock$^a$}

\affiliation{(a) \ C. N. Yang Institute for Theoretical Physics and 
Department of Physics and Astronomy, \\
Stony Brook University, Stony Brook, NY 11794, USA }

\affiliation{(b) \ CP$^3$-Origins, University of Southern Denmark, \\
Campusvej 55, Odense, Denmark}

\begin{abstract}

  We present calculations of certain limits of scheme-independent series
  expansions for the anomalous dimensions of gauge-invariant fermion bilinear
  operators and for the derivative of the beta function at an infrared fixed
  point in SU($N_c$) gauge theories with fermions transforming according to two
  different representations. We first study a theory with $N_f$ fermions in the
  fundamental representation and $N_{f'}$ fermions in the adjoint or 
  symmetric or antisymmetric rank-2 tensor representation, in the limit 
  $N_c \to \infty$, $N_f \to \infty$ with $N_f/N_c$ fixed and finite. We then
  study the $N_c \to \infty$ limit of a theory with fermions in the adjoint and
  rank-2 symmetric or antisymmetric tensor representations.  

\end{abstract}

\maketitle

% =======================================================================

% section 1
\section{Introduction}
\label{intro_section}

In this paper we extend the recent study in Ref. \cite{dexm} on calculations of
scheme-independent series expansions for the anomalous dimensions and the
derivative of the beta function at an infrared fixed point (IRFP) of the
renormalization group in gauge theories with two different fermion
representations.  In Ref. \cite{dexm}, this study was carried out at an IRFP of
an asymptotically free vectorial gauge theory with a general gauge group $G$,
containing massless fermions transforming according to two different
representations of $G$ \cite{fm}. In \cite{dexm} the theory was taken to have
$N_f$ copies (flavors) of Dirac fermions, denoted $f$, in the representation
$R$ of $G$, and $N_{f'}$ copies of fermions, denoted $f'$, in a different
representation $R'$ of $G$. Here we analyze interesting limits of two specific
theories of this type, both of which have the gauge group ${\rm SU}(N_c)$.

In the first type of theory, $R$ is the fundamental representation, denoted
$F$, and $R'$ is any of three types of two-index representations, namely the
adjoint ($Adj$), or the symmetric or antisymmetric rank-2 tensor
representations, denoted $S_2$ and $A_2$, respectively. We call this an $FR'$
theory.  We investigate this $FR'$ theory in the limit
\beqs
& & N_c \to \infty \ , \quad N_F \to \infty  \quad {\rm with} \ 
r \equiv \frac{N_F}{N_c} \ {\rm fixed \ and \ finite}  \cr\cr
& & {\rm and} \ \ \xi(\mu) \equiv \alpha(\mu) N_c \ {\rm is \ a \
finite \ function \ of} \ \mu \ .
\cr\cr
& &
\label{lnn}
\eeqs
We will use the symbol $\lim_{LNN}$ for this limit, where ``LNN'' stands for
``large $N_c$ and $N_F$'' (with the constraints in Eq. (\ref{lnn}) imposed).
This LNN limit, which is often called the 't Hooft-Veneziano limit, has the
simplifying feature that rather than depending on the four quantities $N_c$,
$N_F$, $R'$, and $N_{f'}$, the properties of the theory only depend on three
quantities, namely $r$, $R'$, and $N_{f'}$. A general property that makes the
LNN limit of $FR'$ theories useful is that for large but finite $N_f$ and
$N_c$, the approach to the LNN limit is rapid, because the correction terms to
the limiting expressions vanish like $1/N_c^2$. This was shown in
\cite{lnn,dex,dexl} for theories with fermions in a single representation, and
we report the generalization of this property in the present paper for the
$FR'$ theory. Because of this rapid convergence, one can use calculations of
anomalous dimensions and other physical quantities in the LNN limit with a
given value of $r$ in a unified manner to compare with corresponding
calculations in specific SU($N_c$) theories with various values of $N_f$ and
$N_c$ satisfying $N_f/N_c \simeq r$.

In the second type of theory that we analyze, $R$ and $R'$ are both two-index
representations. We take $R=Adj$ and $R'$ to be $S_2$ or $A_2$, and study the
$N_c \to \infty$ limit of this theory. The leading large-$N_c$ behavior of the
$S_2$ and $A_2$ representations is the same, so that we will often refer to
these jointly as $T_2$, where the symbol $T_2$ stands for rank-2 tensor
representation.  We thus denote this second type of theory as an $AT$ theory,
where $A$ stands for $Adj$ and $T$ for $T_2$.  In contrast to $FR'$ theories,
in which $N_F \to \infty$, in $AT$ theories the requirement of asymptotic
freedom requires that both $N_f=N_{Adj}$ and $N_{f'}=N_{T_2}$ be finite.

In the present paper we shall study the properties of these gauge theories at
an infrared fixed point. We explain the general theoretical background in the
context of an $FR'$ theory and then consider the $AT$ theory. In an $FR'$
theory, the requirement of asymptotic freedom places correlated upper ($u$)
bounds on $r$ and $N_{f'}$, which we denote as $r_u$ and $N_{f',u}$.  Provided
that these bounds are satisfied, the ultraviolet (UV) behavior of the theory
can be well described perturbatively. Then one can explore how the running
gauge coupling $g(\mu)$ changes as a function of the Euclidean energy/momentum
scale $\mu$ where it is measured.  This is described by the beta function,
$\beta(\alpha(\mu)) = d\alpha(\mu)/d\ln\mu$, where $\alpha(\mu) =
g(\mu)^2/(4\pi)$.  (The argument $\mu$ will often be suppressed in the
notation.) Since the theory is asymptotically free, one can calculate the beta
function in a self-consistent manner in the weakly coupled UV region and then
use it to explore the flow (evolution) of the theory from the UV to the IR. For
values of $r$ and $N_{f'}$ near to the above-mentioned upper limits, the beta
function has an IR zero, so the theory flows from the UV to this IR fixed
point. For fixed $N_{f'}$, as $r$ approaches $r_u$ from below, the value of
$\alpha=\alpha_{IR}$ at the IRFP goes to zero. One thus infers that in this
regime, the IR theory is in a deconfined non-Abelian Coulomb phase (NACP)
without any spontaneous chiral symmetry breaking (S$\chi$SB).  Lattice studies
of these types of gauge theories (usually with fermions in a single
representation of the gauge group) with weakly coupled IR fixed points have
supported this conclusion, e.g., by demonstrating the absence of a bilinear
fermion condensate that would signal spontaneous chiral symmetry breaking
\cite{lgtreviews,simons}.  At the IRFP, the resultant theory is scale-invariant
and is deduced to be conformally invariant \cite{scalecon}.  This IR regime is
thus often referred to as the conformal window or regime.  As $r$ and/or
$N_{f'}$ is decreased, the IR coupling $\alpha_{IR}$ increases, and eventually,
for sufficiently small $r$ and $N_{f'}$, the IR theory becomes strongly
coupled, with confinement and S$\chi$SB.  Analogous comments apply to $AT$
theories.

Our scheme-independent calculational framework requires that the IRFP be exact,
which is the case in the conformal regime. Hence we restrict our consideration
to this regime. The properties of the resultant conformal field theory are of
fundamental interest.  Previous works have investigated these properties for a
variety of theories with a general gauge group $G$ and $N_f$ fermions $\psi_i$,
$i=1,...,N_f$ transforming according to a single representation $R$ of $G$,
using perturbative calculations of the anomalous dimension of the operator
$\bar\psi\psi$, denoted $\gamma_{\bar \psi\psi}$, and of the derivative of the
beta function, $d\beta/d\alpha = \beta'$, both evaluated at the IRFP
\cite{lnn}-\cite{dexl}, \cite{bvh}-\cite{dexnote}.  We denote these as
$\gamma_{\bar\psi\psi,IR}$ and $\beta'_{IR}$. Early calculations of this sort
were performed using a perturbative expansion in powers of $\alpha_{IR}$, the
value of $\alpha$ at the IRFP, calculated to the same loop order \cite{bvh,ps}.
Although $\gamma_{\bar\psi\psi,IR}$ and $\beta'_{IR}$ are physical quantities
and hence are independent of the scheme used for regularization and
renormalization, the series expansions for these quantities, calculated to
finite order in powers of $\alpha_{IR}$, are scheme-dependent. This is the same
as in higher-order calculations of scattering cross sections in various quantum
field theories, such as quantum chromodynamics (QCD). However, it is possible
to reexpress the series as expansions in powers of a manifestly
scheme-independent quantity, denoted $\Delta_f$, that approaches zero at the
upper end of the conformal regime \cite{bz}, and for theories with a single
fermion representation, these calculations were carried out to $O(\Delta_f^4)$
for $\gamma_{\bar\psi\psi,IR}$ and to $O(\Delta_f^5)$ for $\beta'_{IR}$
\cite{gtr,gsi,dex,dexl,dexo,pgb}. The calculation of a scheme-independent
series expansion for $\gamma_{\bar\psi\psi,IR}$ to $O(\Delta_f^n)$ requires, as
inputs, conventional series expansions (in powers of $\alpha$) of
$\gamma_{\bar\psi\psi}$ to $n$-loop order and of $\beta$ to $(n+1)$-loop order.
The scheme-independent calculation of $\beta'_{IR}$ to $O(\Delta_f^n)$
requires, as an input, the conventional series calculation of $\beta$ to
$n$-loop order. Thus, the scheme-independent calculations of these quantities
in theories with a single fermion representation have used, as inputs,
conventional four-loop \cite{b4} and five-loop \cite{b5su3,b5} series for
$\beta$ and four-loop series for $\gamma_{\bar\psi\psi}$ \cite{c4}. Recently,
higher-order calculations for gauge theories with multiple fermion
representations were performed \cite{zoller,chetzol}.  Ref. \cite{dexm} used
the results from \cite{zoller,chetzol} to calculate scheme-independent series
for the anomalous dimensions of both types of fermions and for $\beta'_{IR}$ in
a theory with two different types of fermion representations. It is of
considerable interest to use the calculations of Ref. \cite{dexm} to explore
various limits of such theories, and we undertake this work here. 

This paper is organized as follows.  In Section \ref{framework_section} we
discuss the general framework for our work and the LNN limit.  In Sections
\ref{kappa_section} and \ref{betaprime_section} we present our results for
anomalous dimensions of fermion bilinears and for the derivative of the beta
function at the IRFP in the LNN limit of the $FR'$ theory. In Section
\ref{at_section} we present our results for the $N_c \to \infty$ limit of the
$AT$ theory.  Our conclusions are given in Section \ref{conclusion_section}.

% =========================================================================

% section II 
\section{General Framework and LNN Limit of $FR'$ Theory} 
\label{framework_section}

\subsection{Upper Limits on $r$ and $N_{f'}$} 
\label{upper_limits}

In this section we discuss the general theoretical framework for our
calculations. The $N_f$ fermions $f$ in the
representation $R=F$ are denoted as $\psi_i$, $i=1,...,N_f$, and the $N_{f'}$
fermions are denoted as $\chi_j$, $j=1,...,N_{f'}$.  Since the adjoint
representation is self-conjugate, the number of fermions in this
representation, $N_{Adj}$, refers equivalently
to a theory with $N_{Adj}$ Dirac fermions or $2N_{Adj}$ Majorana fermions, so
that in this case, $N_{Adj}$ may take on half-integral physical values. In both
the $FR'$ and $AT$ theories, one may consider a formal extension in which
$N_f$ and/or $N_{f'}$ are generalized to (positive) real numbers, with the
implicit understanding that physical cases occur at integral (and, for the
adjoint representation also half-integral) values.  Indeed, in the LNN limit of
the $FR'$ theory, $N_F$ is replaced by the real variable $r$. 

In general, the property of asymptotic freedom requires that
\beq
N_fT_f + N_{f'}T_{f'} < \frac{11C_A}{4} \ , 
\label{af}
\eeq
where $C_A$, $T_f$, and $T_{f'}$ are group invariants \cite{casimir}. 
In the large-$N_c$ limit, the behaviors of group invariants for the 
$S_2$ and $A_2$ representations are the same to leading order, so, as noted
above, one can consider these representations together as $T_2$. 
For example, $T_{f'} = (N_c \pm 2)/2$ for $f'=S_2, \ A_2$, so 
\beq
\lim_{N_c \to \infty} \frac{T_{S_2}}{N_c} = 
\lim_{N_c \to \infty} \frac{T_{A_2}}{N_c} = \frac{1}{2} \ . 
\label{trace_s2a2}
\eeq
To treat the three representations $Adj, \ S_2, \ A_2$ 
in a unified manner, we define
\beq
\lambda_{R} = \lim_{N_c \to \infty} \frac{T_{R}}{N_c}
\label{eta}
\eeq
so that 
\beq
\lambda_{Adj}=1 
\label{eta_adj}
\eeq
(since $T_{Adj}=N_c$) and
\beq
\lambda_{S_2} = \lambda_{A_2} \equiv \lambda_{T_2} = \frac{1}{2} \ . 
\label{eta_t2}
\eeq
In an $FR'$ theory, for fixed $N_{f'}$, the inequality (\ref{af}) implies the
upper ($u$) limit $N_F < N_{F,u}$, where
\beq
N_{F,u} = \frac{11}{2}N_c - 2N_{f'}T_{f'}
\label{nfu}
\eeq
and for fixed $N_F$, this inequality (\ref{af}) implies the upper bound
$N_{f'} < N_{f',u}$, where 
\beq
N_{f',u} = \frac{11N_c - 2N_F}{4T_{f'}} \ . 
\label{nfprimeu}
\eeq

In the LNN limit of the $FR'$ theory, the inequality (\ref{af}) becomes 
\beq
r + 2\lambda_{f'} N_{f'} < \frac{11}{2} \ . 
\label{af_lnn}
\eeq
For fixed $N_{f'}$, this implies the upper ($u$) limit $r < r_u$, where 
\beq
r_u = \frac{11}{2}-2\lambda_{f'} N_{f'} \ , 
\label{ru}
\eeq
and for fixed $r$, the upper bound on $N_{f'}$ is $N_{f'} < N_{f',u}$, where
\beq
N_{f',u} = \frac{11-2r}{4\lambda_{f'}} \ . 
\label{nfprimeu_lnn}
\eeq
If one envisions a two-dimensional diagram describing the $FR'$ theory with the
horizontal axis being $r$ and the vertical axis being $N_{f'}$ (formally
generalized from the integers to the real numbers), then the inequality
(\ref{af_lnn}) defines a region in the first quadrant bounded by the line
segment $r + 2\lambda_{f'} N_{f'} =0$ extending from the point
$(r,N_{f'})=(0,N_{f',u})$ on the upper left to the the point
$(r,N_{f'})=(r_u,0)$ on the lower right.  This line has slope
\beq
\frac{dN_{f'}}{dr} = -\frac{1}{2\lambda_{f'}} \ . 
\label{upper_line_slope}
\eeq
In order to have a theory with two fermion representations, we exclude
the values $r=0$ and $N_{f'}=0$.  

In the LNN limit of the $FR'$ theory we define the differences 
\beq
\Delta_r = r_u-r = \frac{11}{2} - 2\lambda_{f'}N_{f'} - r
\label{deltar}
\eeq
and
\beqs
\check\Delta_{f'} &=& \lim_{LNN} (N_{f',u}-N_{f'}) \cr\cr
            &=& \frac{11-2r}{4\lambda_{f'}} - N_{f'} \ .
\label{deltafp}
\eeqs
We observe that
\beq
\Delta_r = 2\lambda_{f'}\check\Delta_{f'} \ . 
\label{ddrel}
\eeq
%

% ====================================================================

\subsection{Anomalous Dimensions of Fermion Bilinears and Series Expansions} 
\label{anomdim_section}

We denote the full scaling dimension
of an operator ${\cal O}$ as $D_{\cal O}$ and its free-field value as $D_{{\cal
    O},free}$.  The anomalous dimension of this operator, embodying the effect
of interactions, denoted $\gamma_{\cal O}$, is given by
\beq
D_{\cal O} = D_{{\cal O},free} - \gamma_{\cal O} \ . 
\label{anomdim}
\eeq
The gauge-invariant fermion bilinears considered here are
\beq
\bar f f \equiv \bar\psi\psi = \sum_{j=1}^{N_f} \bar\psi_j \psi_j
\label{psibarpsi}
\eeq
and 
\beq
\bar f' f' \equiv \bar\chi\chi = \sum_{j=1}^{N_{f'}} \bar\chi_j \chi_j \ .
\label{chibarchi}
\eeq
The anomalous dimension of $\bar\psi\psi$ is the same as that of the
bilinear $\sum_{j,k=1}^{N_f} \bar\psi_j {\cal T}_a \psi_k$,
where ${\cal T}_a$ is a generator of the Lie algebra of SU($N_f$)
\cite{gracey_gammatensor}, so we use the same symbol $\gamma_{\bar\psi\psi}$ 
for both. The same remark holds for $\gamma_{\bar\chi\chi}$.

Because $\alpha_{IR} \to 0$ at the upper end of the conformal regime, a series
expansion for an anomalous dimension of a fermion bilinear or for $\beta'_{IR}$
can be reexpressed as a series expansion in powers of the manifestly
scheme-independent quantities $\Delta_r$ and/or $\Delta_{f'}$.
For finite $N_c$ and $N_f = N_F$, the scheme-independent series expansion of 
$\gamma_{\bar\psi\psi,IR}$ and $\gamma_{\bar \chi\chi,IR}$ are 
\beq
\gamma_{\bar\psi\psi,IR} = \sum_{j=1}^\infty \kappa^{(f)}_j \, \Delta_f^j
\label{gamma_ir_Deltaseries}
\eeq
and
\beq
\gamma_{\bar\chi\chi,IR} = \sum_{j=1}^\infty \kappa^{(f')}_j \, \Delta_{f'}^j
\ . 
\label{gamma_ir_Delta2series}
\eeq
In the LNN limit of the $FR'$ theory, $\kappa^{(F)}_j \propto N_c^{-j}$ and 
$\kappa^{(f')}_j \propto N_c^{0}$, so one defines a rescaled 
$\kappa^{(F)}_j$ coefficient as 
\beq
\hat\kappa^{(F)}_j = \lim_{LNN} N_c^j \, \kappa^{(F)}_j \ ,
\label{kappajhat}
\eeq
and one defines the limit 
\beq
\bar\kappa^{(f')}_j = \lim_{LNN} \kappa^{(f')}_j \ . 
\label{kappajbar}
\eeq

The scheme-independent series expansions for the anomalous dimensions of the
gauge-invariant fermion bilinear operators in the $FR'$ theory, evaluated at
the IRFP, namely $\gamma_{\bar\psi\psi,IR}$ and $\gamma_{\bar\chi\chi,IR}$, are
then as follows, in the LNN limit:
\beq
\gamma_{\bar\psi\psi ,IR} = \sum_{j=1}^\infty \hat\kappa^{(F)}_j \, \Delta_r^j
\label{gamma_ir_Deltaseries_lnn}
\eeq
and 
\beq
\gamma_{\bar\chi\chi,IR} = \sum_{j=1}^\infty \bar\kappa^{(f')}_j \, 
\check\Delta_{f'}^j \ . 
\label{gamma_ir_Delta2series_lnn}
\eeq
We denote the truncations of these series to the power $p$ of the respective
expansion variable $\Delta_r$ or $\Delta_{f'}$ as 
$\gamma_{\bar\psi\psi,IR,\Delta_r^p}$ and 
$\gamma_{\bar\chi\chi,IR,\check\Delta_{f'}^p}$, respectively. A corresponding
discussion of scheme-independent series expansions of anomalous dimensions of
bilinear fermion operators in the $AT$ theory is given in Section
\ref{at_section}. 

% ========================================================================

\subsection{Series for $\beta'_{IR}$ }
\label{betaprime}

The series expansion of $\beta$ in powers of the squared gauge coupling is
\beq
\beta = -2\alpha \sum_{\ell=1}^\infty b_\ell \, a^\ell \ , 
\label{beta}
\eeq
where $a=\alpha/(4\pi)$ and $b_\ell$ is the $\ell$-loop coefficient. As was
specified in Eq. (\ref{lnn}), the product $\xi = N_c \alpha$ is fixed in the 
LNN limit. Hence, one deals with the rescaled beta 
function that is finite in this LNN limit, namely 
\beq
\beta_\xi = \frac{d\xi}{d\ln\mu} = \lim_{LNN} N_c \beta \ . 
\label{betaxi}
\eeq
This has the series expansion 
\beq
\beta_\xi \equiv \frac{d\xi}{dt}
= -2\xi \sum_{\ell=1}^\infty \hat b_\ell x^\ell \ ,
\label{betaxiseries}
\eeq
where $x=\xi/(4\pi)$ and 
\beq
\hat b_\ell = \lim_{LNN} \frac{b_\ell}{N_c^\ell} \ .
\label{bellhat}
\eeq

Because the derivative $d\beta_\xi/d\xi$ satisfies 
\beq
\frac{d \beta_\xi}{d\xi} = \frac{d\beta}{d\alpha} \equiv \beta' \ ,
\label{dbetarelation}
\eeq
a consequence is that $\beta'$ is finite in the LNN limit (\ref{lnn}).  There
are two equivalent scheme-independent series expansions of the derivative
$\beta'_{IR}$. One can take $N_{f'}$ as fixed and $N_f$ as variable and write
the series as an expansion in powers of $\Delta_F$:
\beq
\beta'_{IR} = \sum_{j=2}^\infty d_j \, \Delta_F^j \ .  
\label{betaprime_ir_Deltaseries}
\eeq
Equivalently, one may take $N_f$ as fixed and $N_{f'}$ as variable, and express
the series as an expansion in powers of $\Delta_{f'}$,
as
\beq 
\beta'_{IR} = \sum_{j=2}^\infty \tilde d_j \, \Delta_{f'}^j \ .
\label{betaprime_ir_Deltaseries2}
\eeq
Note that $d_1 = \tilde d_1 = 0$ for all $G$ and fermion representations. 
In the LNN limit, $d_j \propto N_c^{-j}$ and $\tilde d_j \propto N_c^0$, so we
define rescaled coefficients 
\beq
\hat d_j = \lim_{LNN} N_c^j \, d_j
\label{djhat_lnn}
\eeq
and
\beq
\bar d_j = \lim_{LNN} \tilde d_j \ . 
\label{djbar_lnn}
\eeq
The scheme-independent expansions for $\beta'$ then take the form
\beq
\beta'_{IR} = \sum_{j=2}^\infty \hat d_j \, \Delta_r^j 
\label{betaprime_ir_Deltaseries_lnn}
\eeq
and
\beq
\beta'_{IR} = \sum_{j=2}^\infty \bar d_j \, \check\Delta_{f'}^j \ . 
\label{betaprime_ir_Deltaseries2_lnn}
\eeq
We denote the truncation of the series expansion 
(\ref{betaprime_ir_Deltaseries_lnn}) to maximal power $\Delta_r^p$ as
$\beta'_{IR,\Delta_r^p}$ and the trunction of the series expansion 
(\ref{betaprime_ir_Deltaseries2_lnn}) to maximal power $\check\Delta_{f'}^p$ as
$\beta'_{IR,\check\Delta_{f'}^p}$. 

% ========================================================================

\subsection{Relevant Ranges of $(r,N_{f'})$} 
\label{ranges}

Our scheme-independent calculations require that the IRFP be exact. This
condition is satisfied in the conformal regime but not in the QCD-like regime
with spontaneous chiral symmetry breaking.  The upper boundary of this regime
is known precisely and is given by the inequality (\ref{af_lnn}).  The lower
boundary of the conformal regime is not known precisely and has been the
subject of intensive lattice studies \cite{lgtreviews,simons}, particularly for
simpler theories with fermions in a single representation.  Further lattice
studies could be carried out for theories with multiple fermion
representations. For instance, a study has been carried out of an SU(4) gauge
theory with $N_f=2$ Dirac fermions in the fundamental representation and
$N_{f'}=2$ Dirac fermions in the (self-conjugate) antisymmetric rank-2 tensor
representation \cite{su4lgt1,su4lgt2}, concluding that this theory is in the
phase with chiral symmetry breaking for both types of fermions.

For our present purposes, it will be sufficient to have a rough guide to this
lower boundary of the conformal regime, which is provided by the condition that
the two-loop (rescaled) beta function should have an IR zero. This condition is
satisfied if the two-loop coefficient in the beta function has a sign opposite
to that of the one-loop coefficient, i.e., if the inequality
\beq
13r + 32\lambda_{f'}N_{f'} - 34 > 0
\label{b2inequality}
\eeq
is satisfied.  For a given $N_{f'}$, this yields a lower ($\ell$) 
bound on $r$, namely $r > r_\ell$, where 
\beq
r_{\ell} = \frac{34-32\lambda_{f'}N_{f'}}{13} \ , 
\label{rell}
\eeq
and for a given $r$ a lower bound on $N_{f'}$, namely $N_{f'} > N_{f',\ell}$,
where 
\beq
N_{f',\ell} = \frac{34-13r}{32\lambda_{f'}} \ . 
\label{nf2ell}
\eeq
We denote the set of values of $r$ and $N_{f'}$ which satisfy the asymptotic
freedom constraint and the inequality (\ref{b2inequality}) as $I_{IRZ}$, where
the subscript $IRZ$ refers to the condition that the two-loop beta function has
an IR zero. Henceforth, we assume that if $N_{f'}$ is fixed, then $r \in
I_{IRZ}$ and if $r$ is fixed, then $N_{f'} \in I_{IRZ}$.  The upper end of the
IRZ region is defined the asymptotic freedom constraint (\ref{af}), while the
lower end is defined by the line segment $13r+32\lambda_{f'}N_{f'}-34=0$ in
the first quadrant.  This line segment extends from the point
$(0,17/(16\lambda_{f'})$ at the upper left down to the point $(34/13,0)$ on the
lower right, with slope
\beq
\frac{dN_{f'}}{dr} = -\frac{13}{32\lambda_{f'}} \ . 
\label{lower_line_slope}
\eeq
In Table \ref{rinterval} we list the values of $r_\ell$ and $r_u$ for a range
of values of $N_{Adj}$ and $N_{T_2}$.  For a given $r$, the condition of
asymptotic freedom sets the upper bound $N_{f',u}$ on $N_{f'}$, and this
determines the values of $N_{f'}$ given in Table \ref{rinterval} for $R'=Adj$
and $R'=T_2$.

Provided that, $r$ and $N_{f'}$ satisfy the asymptotic freedom constraint
(\ref{af}) and  lie in the set of values $I_{IRZ}$, ed by the asymptotic
freedom condition (\ref{af_lnn}), the ratio $r$ is in the interval $I_{IRZ}$,
the IR zero in the rescaled two-loop beta function of the $FR'$ theory occurs at
\beq
\xi_{IR,2\ell} = \frac{4\pi \Big [11-2(r+2\lambda_{f'}N_{f'}) \Big ] }
{13r+32\lambda_{f'}N_{f'}-34} \ ,
\label{xiir_2loop}
\eeq
where $\xi$ was defined in (\ref{lnn}).  For a given $R_{f'}$ and $N_{f'}$, as
$r \nearrow r_u$, this IR zero, and more generally the $n$-loop IR zero of
$\beta_\xi$, vanishes. Similarly, for a given $R_{f'}$ and $r$, as $N_{f'}
\nearrow N_{f',u}$ (with $N_{f'}$ generalized to a real number, as above), the
IR zero of the beta function vanishes.

% ========================================================================

\section{Anomalous Dimensions of Fermion Bilinear Operators in $FR'$ Theory}
\label{kappa_section}

In the LNN limit of the $FR'$ theory, from \cite{dexm} we calculate the
following results for the coefficients in the scheme-independent expansions of
$\gamma_{\bar\psi\psi,IR}$ and $\gamma_{\bar\chi\chi,IR}$, where $f \equiv 
\psi$ is in the $F$ representation and $f' \equiv \chi$ is in the 
$R'$ representation: 
\beq
\hat \kappa^{(F)}_1 = \frac{4}{25 +4\lambda_{f'}N_{f'}} 
\label{kappa1hat_fund}
\eeq
\beq
\hat \kappa^{(F)}_2 = \frac{4(147+40\lambda_{f'}N_{f'})}
                           {(25+4\lambda_{f'}N_{f'})^3}
\label{kappa2hat_fund}
\eeq
\beq
\hat\kappa^{(F)}_3 = \frac{2^3[274243+135848\lambda_{f'}N_{f'} + 
22048(\lambda_{f'}N_{f'})^2]}{3^3(25+4\lambda_{f'}N_{f'})^5}
\label{kappa3hat_fund}
\eeq
\beq
\bar \kappa^{(f')}_1 = \frac{2^3\lambda_{f'}}{18-r}
\label{kappa1bar_fprime}
\eeq
\beq
\bar \kappa^{(f')}_2 = \frac{2^2(1023-74r)\lambda_{f'}^2}{3(18-r)^3} 
\label{kappa2bar_fprime}
\eeq
and 
\beq
\bar \kappa^{(f')}_3 = \frac{2^2(1670571-242208r+9184r^2)\lambda_{f'}^3}
                            {3^3(18-r)^5} \ . 
\label{kappa3bar_fprime}
\eeq
Here and below, we indicate the simple factorizations of numbers appearing in
denominators. (The numbers in the numerators do not, in general, have such
simple factorizations; for example, in $\hat \kappa^{(F)}_3$, the number 274243
is prime.) We record values of the $\hat\kappa^{(F)}_j$ as functions of $r$ in
Table \ref{kappahat_values}.  For the illustrative case $R'=Adj$, we also list
values of $\bar\kappa^{(f')}_j = \bar\kappa^{(Adj)}_j$ in Table
\ref{kappabar_values}.  Generalizing the earlier findings for theories with
fermions in a single representation \cite{lnn,dex,dexl}, we find that the
corrections to these limits (\ref{kappa1hat_fund})-(\ref{kappa3bar_fprime})
vanish like $1/N_c^2$ as $N_c \to \infty$.

An important result that was found in previous work \cite{gsi}-\cite{dexo} was
that for a theory with a single representation, $\kappa^{(f)}_1$ and
$\kappa^{(f)}_2$ are manifestly positive, and for all of the specific gauge
groups and fermion representations that were considered, $\kappa^{(f)}_3$ and
$\kappa^{(f)}_4$ are also positive. This property implied several monotonicity
relations for the calculation of $\gamma_{\bar\psi\psi}$ to maximal power
$\Delta_f^p$, denoted $\gamma_{\bar\psi\psi,\Delta_f^p}$, namely that (for all
$p$ calculated there, i.e., $1 \le p \le 4$), (i) for
fixed $p$, $\gamma_{\bar\psi\psi,\Delta_f^p}$ is a monotonically increasing
function of $\Delta_f$, i.e., a monotonically increasing function of decreasing
$N_f$, and (ii) for fixed $N_f$, $\gamma_{\bar\psi\psi,\Delta_f^p}$ is a
monotonically increasing function of the maximal power $p$.

This positivity question was explored further in \cite{dexm}, and it was shown
that both $\hat\kappa^{(F)}_j$ and $\bar\kappa^{(f')}_j$ are positive for all
of the orders that were calculated, namely $j=1, \ 2, \ 3$. This then implied
the same monotonicity theorems as mentioned above for all of the truncation
orders calculated in \cite{dexm}, namely $1 \le p \le 3$. Here we extend this
analysis to the LNN limit of an $FR'$ theory. We again find that
$\hat\kappa^{(F)}_j$ and $\bar\kappa^{(f')}_j$ are positive for $j=1, \ 2, \ 3$
and for all $r$ and values of $N_{f'}$ considered here, in particular, all of
the values satisfying the conditions (\ref{af}) and (\ref{b2inequality}) for
all two-index representations for $f'$.  This implies four monotonicity
relations for $\gamma_{\bar\psi\psi,\Delta_r^p}$ and
$\gamma_{\bar\chi\chi,\Delta_{f'}^p}$ (in the conformal regime where our
calculations apply), which are the generalizations of the above-mentioned two
relations to the $FR'$ theory.  We list these as the first four relations
below. One may also investigate how $\gamma_{\bar\psi\psi,\Delta_r^p}$ depends
on $N_{f'}$ and how $\gamma_{\bar\chi\chi,\Delta_{f'}^p}$ depends on $r$. As an
input to this determination, we find that the coefficients $\hat\kappa^{(F)}_j$
are monotonically decreasing functions of $N_{f'}$.  Our monotonicity relations
are then as follows:

\begin{enumerate}

\item For fixed $p$ and $N_{f'}$, $\gamma_{\bar\psi\psi,\Delta_r^p}$ is a 
monotonically increasing function of $\Delta_r$, and hence, given the
expression for $\Delta_r$ in Eq. (\ref{deltar}), this anomalous dimension 
decreases monotonically as $r$ increases (and vanishes as $r$ approaches its
upper limit, $r_u$).

\item For fixed $p$ and $r$, $\gamma_{\bar\chi\chi,\Delta_{f'}^p}$ is a
  monotonically increasing function of $\Delta_{f'}$, i.e., this anomalous
  dimension decreases monotonically with increasing $N_{f'}$ (and vanishes as 
$N_{f'}$, formally generalized from integers to real numbers, 
approaches its upper limit, $N_{f',u}$). 

\item For fixed $r$ and $N_{f'}$, $\gamma_{\bar\psi\psi,\Delta_{f'}^p}$ is a
monotonically increasing function of the maximal power $p$.

\item For fixed $r$ and $N_{f'}$, $\gamma_{\bar\chi\chi,\Delta_{f'}^p}$ is a
monotonically increasing function of the maximal power $p$.

\item Because of the positivity of $\kappa^{(F)}_j$, combined with the property
  that the $\kappa^{(F)}_j$ are decreasing functions of $N_{f'}$ and the
  property that $\Delta_r$ is a decreasing function of both $r$ and $N_{f'}$,
  it follows that for fixed $p$ and $r$, $\gamma_{\bar\psi\psi,\Delta_r^p}$ is a
  monotonically decreasing function of $N_{f'}$ and for fixed $p$ and $N_{f'}$,
  $\gamma _{\bar\psi\psi,\Delta_r^p}$ is a decreasing function of $r$.

\end{enumerate}

Although we find that the coefficients $\kappa^{(f')}_j$ are monotonically 
increasing functions of $r$, this trend is outweighed by the property that 
$\Delta_{f'}$ is a monotonically decreasing function of both $r$ and $N_{f'}$,
so that for fixed $p$ and $r$, $\gamma_{\bar\chi\chi,\Delta_{f'}^p}$ is a
monotonically decreasing function of $N_{f'}$ as $N_{f'} \nearrow N_{f',u}$ and
for fixed $p$ and $N_{f'}$, $\gamma_{\bar\chi\chi,\Delta_{f'}^p}$ is a
monotonically decreasing function of $r$ as $r \nearrow r_u$. In both of these
limits, $\gamma_{\bar\chi\chi,\Delta_{f'}^p} \to 0$. 

The first, second, and fifth relations, as well as the relation just given, 
can be understood physically as a consequence of the fact that these 
anomalous dimensions result from the gauge interactions, and (a) 
for fixed $N_{f'}$, increasing $r$ to $r_u$ or (b)
for fixed $r$, increasing $N_{f'}$ (formally generalized from integers to real
numbers) to $N_{f',u}$ leads to a vanishing value of $\alpha_{IR}$.  Hence, in
these limits, since $\alpha_{IR} \to 0$, so do the anomalous dimensions of
these fermion bilinears. 

% =====================================================================

\begin{figure}
  \begin{center}
    \includegraphics[height=8cm]{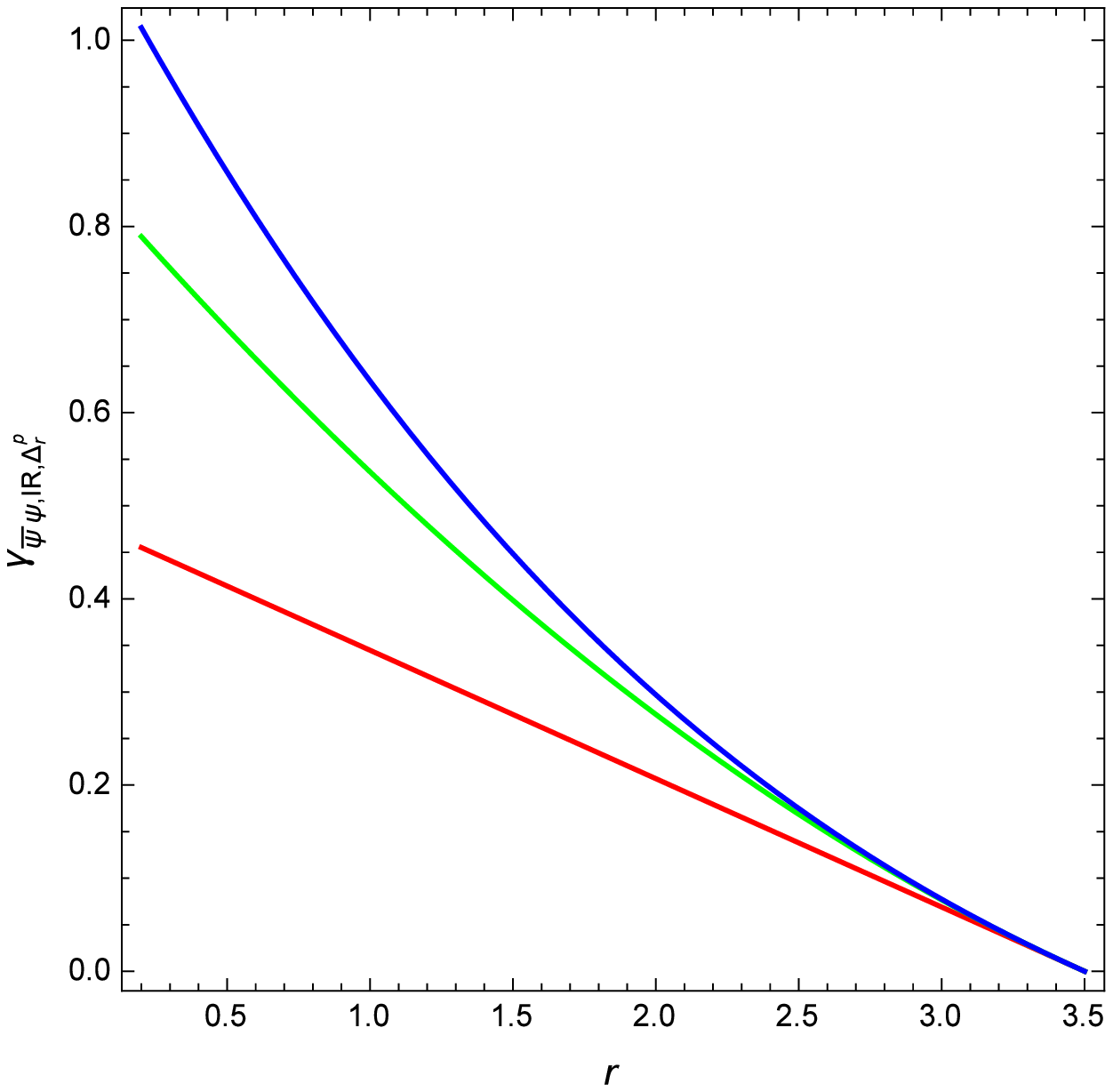}
  \end{center}
\caption{Plot of $\gamma_{\bar\psi\psi,IR,F,\Delta_r^p}$ 
as a function of $r$ in the $FR'$ theory with $R'=Adj$ and
$N_{f'} \equiv N_{Adj}=1$. 
From bottom to top, the curves (with colors online) refer to
$\gamma_{\bar\psi\psi,IR,F,\Delta_r}$ (red),
$\gamma_{\bar\psi\psi,IR,F,\Delta_r^2}$ (green) and
$\gamma_{\bar\psi\psi,IR,F,\Delta_r^3}$ (blue).}
\label{gamma_psibarpsi_Nadj1_figure}
\end{figure}

\begin{figure}
  \begin{center}
    \includegraphics[height=8cm]{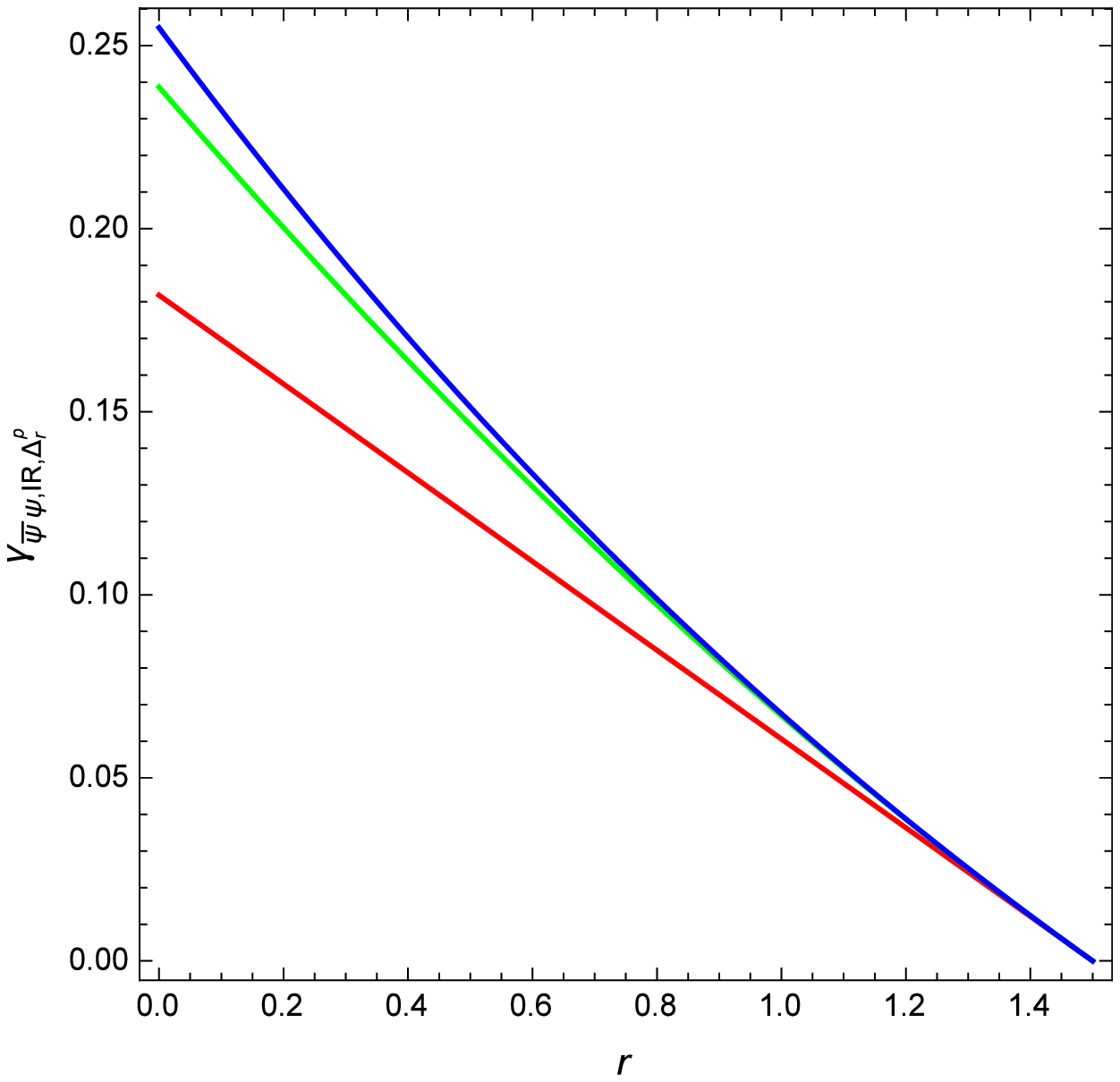}
  \end{center}
\caption{Plot of $\gamma_{\bar\psi\psi,IR,F,\Delta_r^p}$
as a function of $r$ in the $FR'$ theory 
for the case $R'=Adj$ and $N_{f'} \equiv N_{Adj}=2$. 
From bottom to top, the curves (with colors online) refer to
$\gamma_{\bar\psi\psi,IR,F,\Delta_r}$ (red),
$\gamma_{\bar\psi\psi,IR,F,\Delta_r^2}$ (green) and
$\gamma_{\bar\psi\psi,IR,F,\Delta_r^3}$ (blue).}
\label{gamma_psibarpsi_Nadj2_figure}
\end{figure}

\begin{figure}
  \begin{center}
    \includegraphics[height=8cm]{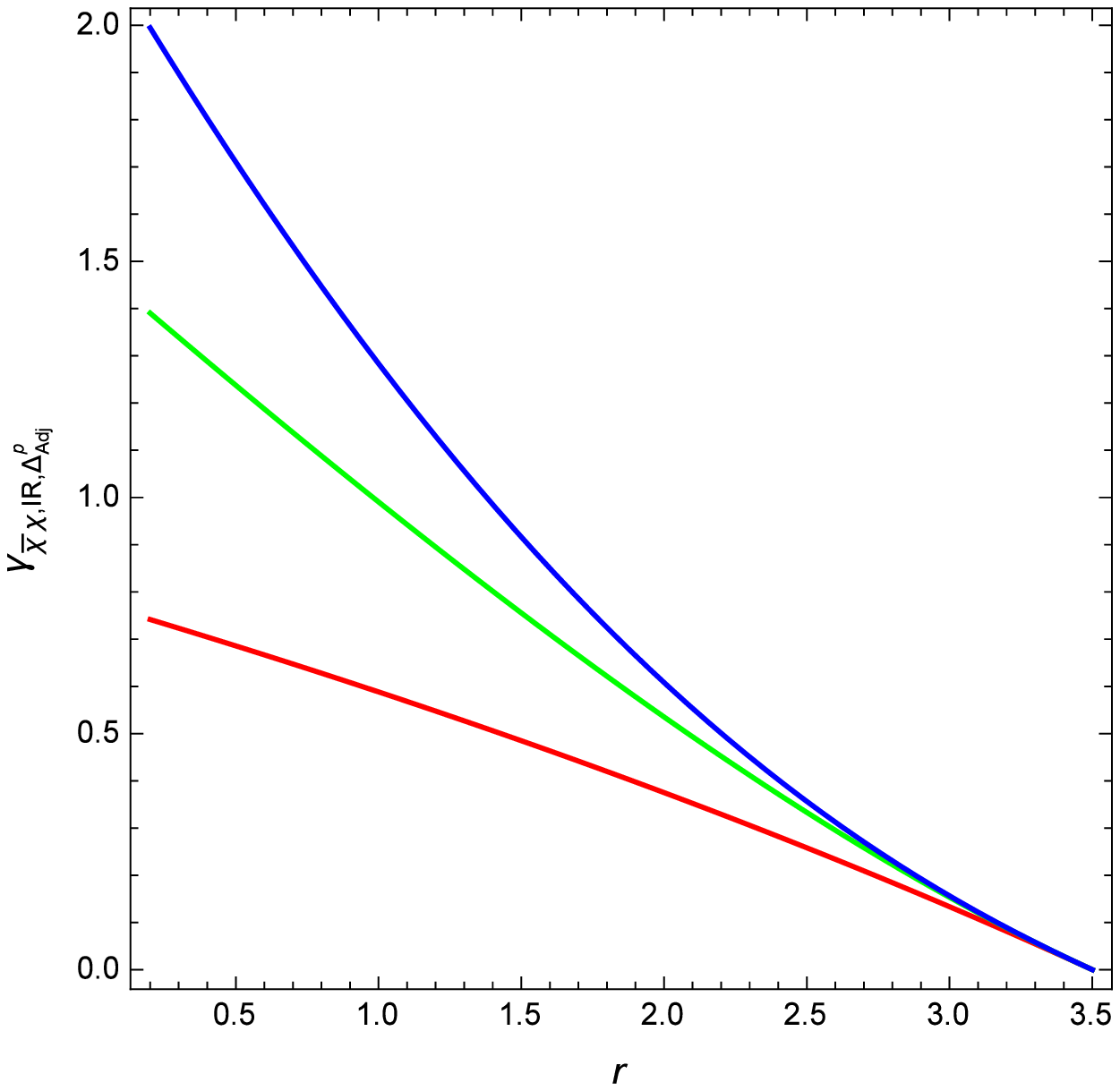}
  \end{center}
\caption{Plot of $\gamma_{\bar\chi\chi,IR,F,\check\Delta_{Adj}^p}$
as a function of $r$ in the $FR'$ theory 
 for the case $R'=Adj$ and $N_{f'} \equiv N_{Adj}=1$. 
From bottom to top, the curves (with colors online) refer to
$\gamma_{\bar\chi\chi,IR,F,\check\Delta_{Adj}}$ (red),
$\gamma_{\bar\chi\chi,IR,F,\check\Delta_{Adj}^2}$ (green) and
$\gamma_{\bar\chi\chi,IR,F,\check\Delta_{Adj}^3}$ (blue).}
\label{gamma_chibarchi_Nadj1_figure}
\end{figure}

\begin{figure}
  \begin{center}
    \includegraphics[height=8cm]{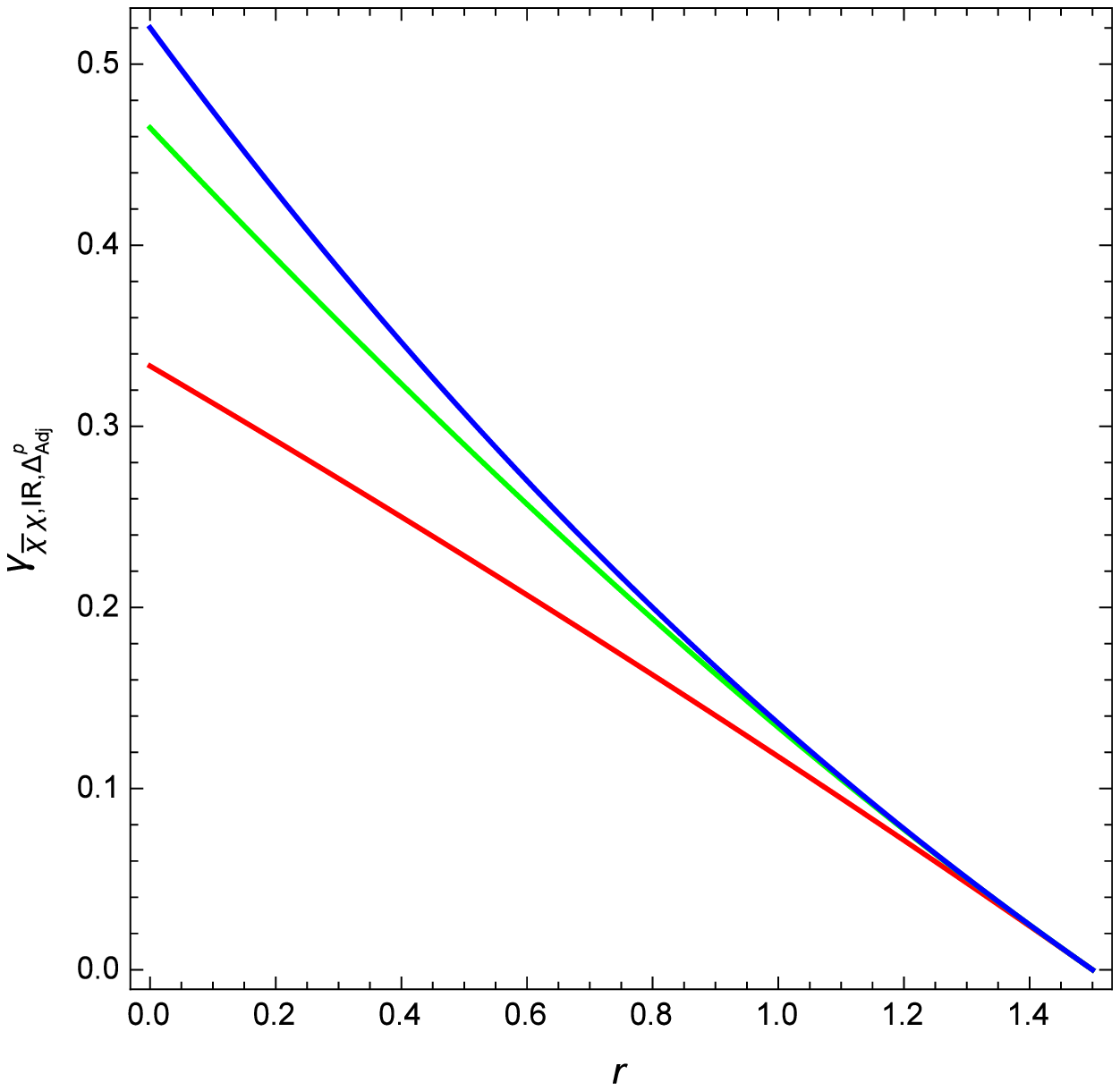}
  \end{center}
\caption{Plot of $\gamma_{\bar\chi\chi,IR,F,\check\Delta_{Adj}^p}$
as a function of $r$ in the $FR'$ theory with 
$R'=Adj$ and $N_{f'} \equiv N_{Adj}=2$. 
From bottom to top, the curves (with colors online) refer to
$\gamma_{\bar\chi\chi,IR,F,\check\Delta_{Adj}}$ (red),
$\gamma_{\bar\chi\chi,IR,F,\check\Delta_{Adj}^2}$ (green) and
$\gamma_{\bar\chi\chi,IR,F,\check\Delta_{Adj}^3}$ (blue).}
\label{gamma_chibarchi_Nadj2_figure}
\end{figure}

We next insert these calculated coefficients $\hat\kappa^{(F)}_j$ and
$\bar\kappa^{(Adj)}_j$ into the general scheme-independent expansions
(\ref{gamma_ir_Deltaseries}) for $f$ with $R=F$ and
(\ref{gamma_ir_Delta2series}) for $f'$. We show the results for
$\gamma_{\bar\psi\psi,IR,\Delta_r^p}$ and 
$\gamma_{\bar\chi\chi,IR,\check\Delta_{Adj}^p}$ in Tables
\ref{gamma_fund_lnn_Nf2_eq_1}-\ref{gamma_adj_lnn_Nf2_eq_2} for two
illustrative cases, namely $R_{f'}=Adj$, $N_{f'} \equiv N_{Adj}=1$, and
$N_{Adj}=2$.  We present plots of $\gamma_{\bar\psi\psi,IR,\Delta_r^p}$ and
$\gamma_{\bar\chi\chi,IR,\check\Delta_r^p}$ with $1 \le p \le 3$ for these two
theories in Figs. \ref{gamma_psibarpsi_Nadj1_figure}-
\ref{gamma_chibarchi_Nadj2_figure}.

It is of interest to compare the values of
$\gamma_{\bar\psi\psi,IR,\Delta_r^p}$ and
$\gamma_{\bar\chi\chi,IR,\check\Delta_{Adj}r^p}$ for $r=10/3$ 
with the results in the
SU(3) theory with $N_F=10$, $R_{f'}=Adj$, and $N_{f'}=1$ given, respectively,
in Tables V and VI of \cite{dexm}.  For that SU(3) theory one has $r=10/3$. In
that theory, for the successive truncations to progressively high order for the
scheme-independent series for $\gamma_{\bar\psi\psi,IR}$ we obtained
$\gamma_{\bar\psi\psi,IR,\Delta_F}=0.0210$,
$\gamma_{\bar\psi\psi,IR,\Delta_F^2}=0.0218$, and
$\gamma_{\bar\psi\psi,IR,\Delta_F^3}=0.0218$, as listed in Table V of
\cite{dexm}. The LNN values that we have listed for $r=10/3$ in Table
\ref{gamma_fund_lnn_Nf2_eq_1} are close to these for each order of truncation.
In the above-mentioned SU(3) theory with $N_F=10$, $R_{f'}=Adj$, and $N_{f'}=1$
we calculated $\gamma_{\bar\chi\chi,IR,\Delta_F}=0.0.0466$,
$\gamma_{\bar\chi\chi,IR,\Delta_F^2}=0.0490$, and
$\gamma_{\bar\chi\chi,IR,\Delta_F^3}=0.0491$, as listed in Table V of
\cite{dexm}. Again, the LNN values that we have listed for $r=10/3$ in Table
\ref{gamma_adj_lnn_Nf2_eq_1} are close to these for each order of
truncation. This is in agreement with our general result that for even moderate
values of $N_c$ and $N_F$ with $N_F/N_c=r$, and a given $R_{f'}$ and $N_{f'}$,
the resulting anomalous dimensions are approximately given by the LNN limit 
with these values of $r$, $R_{f'}$, and $N_{f'}$, since correction terms to the
LNN limit vanish rapidly, like $1/N_c^2$.  As mentioned above, this was shown
earlier for theories with fermions in a single representation of the gauge
group, and our results here generalize this property to the LNN limit of the
$FR'$ theory. 

% =======================================================================

\section{LNN Limit for Scheme-Independent Beta Function Coefficients in
$FR'$ Theory}
\label{betaprime_section}

In the LNN limit, from \cite{dexm}, we calculate 
\beq
\hat d_2 = \frac{2^4}{3^2(25 + 4 \lambda_{f'}N_{f'})}
\label{d2hat}
\eeq
\beq
\hat d_3 = \frac{2^5 \cdot 13}{3^3(25+4\lambda_{f'}N_{f'})^2}
\label{d3hat}
\eeq
and 
\begin{widetext}
\beqs
\hat d_4 &=& \frac{2^4}{3^5(25+4\lambda_{f'}N_{f'})^5} \, 
\Bigg [ 2(183391-330000\zeta_3)+2^4(1151+1800\zeta_3)\lambda_{f'}N_{f'} \cr\cr
&+&2^4(-3161 + 3744\zeta_3)(\lambda_{f'}N_{f'})^2 + 
2^8(-23+24\zeta_3)(\lambda_{f'}N_{f'})^3 \Bigg ] \ , 
\eeqs
\end{widetext}
where $\zeta_s = \sum_{n=1}^\infty n^{-s}$ is the Riemann zeta function. 
For the $\bar d_j$, we find
\beq
\bar d_2 = \frac{2^5\lambda_{f'}^2}{3^2(18-r)}
\label{d2bar_adj}
\eeq
\beq
\bar d_3 = \frac{2^{10}\lambda_{f'}^3}{3^3(18-r)^2}
\label{d3bar_adj}
\eeq
and 
\begin{widetext}
\beq
\bar d_4 = \frac{2^3\lambda_{f'}^4}{3^5(18-r)^5} \, \Bigg [ 3^3 \cdot 46871 
+ 2^2 \cdot 3^4(143-768\zeta_3)r
+ 2^2(-5153+6912\zeta_3)r^2  +2^5(23-24\zeta_3)r^3 \Bigg ] \ . 
\label{d4bar_adj}
\eeq
\end{widetext}
We then substitute these results for $\hat d_j$ and $\bar d_j$ in Eqs.
(\ref{betaprime_ir_Deltaseries_lnn}) and (\ref{betaprime_ir_Deltaseries2_lnn})
with $f'=Adj$, respectively, to obtain the series expansions for $\beta'_{IR}$
in the theory with $R=F$ and $R'=Adj$.

\begin{figure}
  \begin{center}
    \includegraphics[height=8cm]{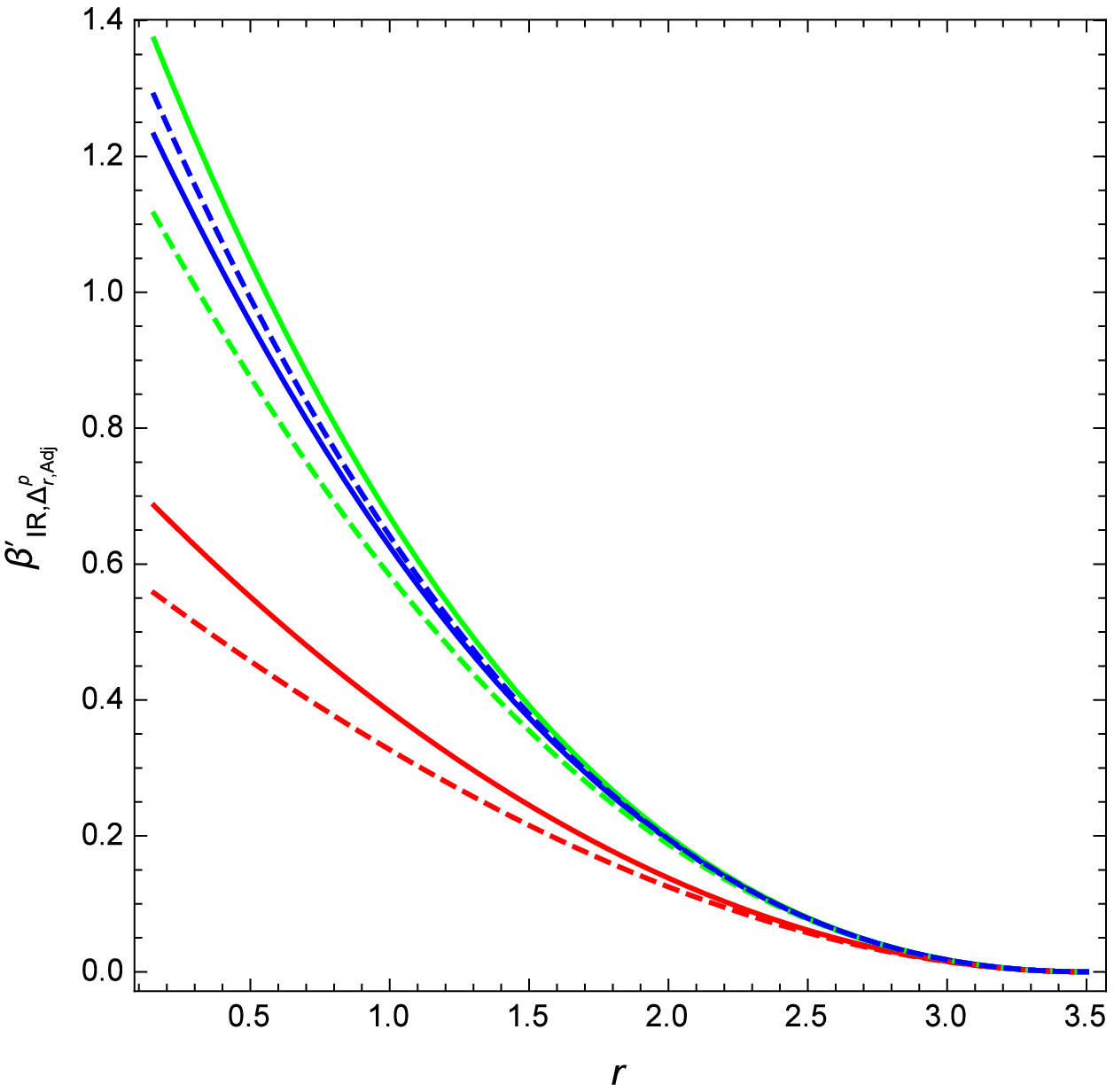}
  \end{center}
\caption{Plots of $\beta'_{IR}$, as calculated with the expansion 
(\ref{betaprime_ir_Deltaseries}) (solid curves) and 
(\ref{betaprime_ir_Deltaseries2}) (dashed curves) with $p=2, \ 3, \ 4$, 
as a function of $r$ in the $FR'$ theory with $R'=Adj$ and 
$N_{f'} \equiv N_{Adj}=1$. 
The curves (with colors online) are as follows: 
$\beta'_{IR,\Delta_r^2}$ (solid red),
$\beta'_{IR,\check\Delta_{Adj}^2}$ (dashed red), 
$\beta'_{IR,\Delta_r^3}$ (solid green),
$\beta'_{IR,\check\Delta_{Adj}^3}$ (dashed green), 
$\beta'_{IR,\Delta_r^4}$ (solid blue), and
$\beta'_{IR,\check\Delta_{Adj}^4}$ (dashed blue).}
\label{betaprime_Nadj1_figure}
\end{figure}

\begin{figure}
  \begin{center}
    \includegraphics[height=8cm]{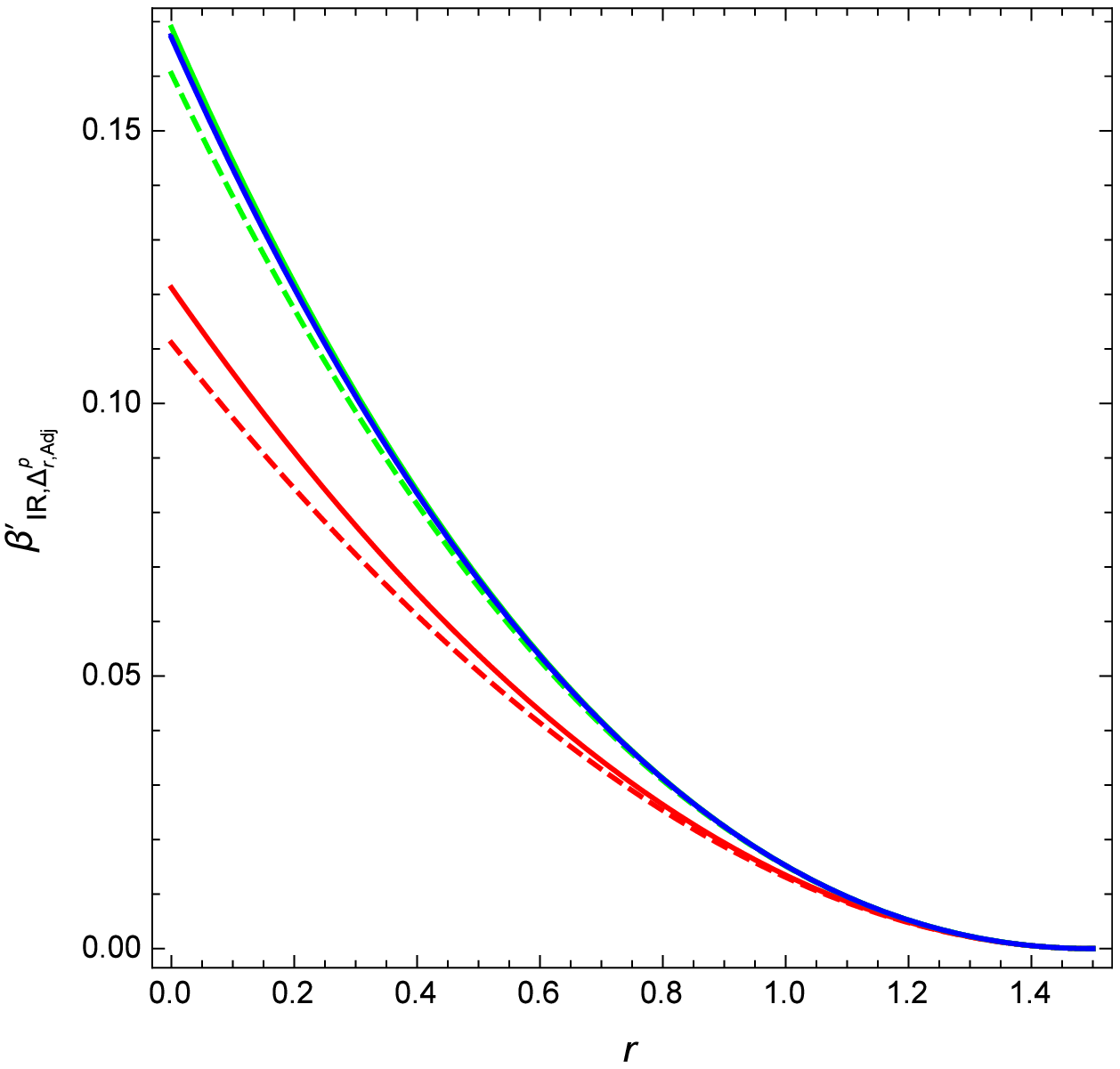}
  \end{center}
\caption{Plots of $\beta'_{IR}$ as calculated with the expansion 
(\ref{betaprime_ir_Deltaseries}) (solid curves) and 
(\ref{betaprime_ir_Deltaseries2}) (dashed curves) with $p=2, \ 3, \ 4$, 
as a function of $r$ in the $FR'$ theory with 
$R'=Adj$ and $N_{f'} \equiv N_{Adj}=2$. 
The curves (with colors online) are as follows: 
$\beta'_{IR,\Delta_r^2}$ (solid red),
$\beta'_{IR,\check\Delta_{Adj}^2}$ (dashed red), 
$\beta'_{IR,\Delta_r^3}$ (solid green),
$\beta'_{IR,\check\Delta_{Adj}^3}$ (dashed green), 
$\beta'_{IR,\Delta_r^4}$ (solid blue), and
$\beta'_{IR,\check\Delta_{Adj}^4}$ (dashed blue).}
\label{betaprime_Nadj2_figure}
\end{figure}

We present our results using the two equivalent scheme-independent series
expansions for $\beta'_{IR}$ in Tables \ref{betaprime_lnn_values_Nadj_eq_1} and
\ref{betaprime_lnn_values_Nadj_eq_2} for our illustrative $FR'$ theories in the
LNN limit with $R_{f'} \equiv R'=Adj$ and $N_{Adj}=1$ and $N_{Adj}=2$,
respectively, as a function of $r$. From left to right in these tables, the
columns list $r$ and the successively higher truncations of the series
expansions, namely $\beta'_{IR,\Delta_r^2}$,
$\beta'_{IR,\check\Delta_{Adj}^2}$, $\beta'_{IR,\Delta_r^3}$,
$\beta'_{IR,\check\Delta_{Adj}^3}$, $\beta'_{IR,\Delta_r^4}$, and
$\beta'_{IR,\check\Delta_{Adj}^4}$.  We see that for a given order $p$ of
truncation, the alternate series expansion values $\beta'_{IR,\Delta_r^p}$ and
$\beta'_{IR,\Delta_{Adj}^p}$ agree reasonably well with each other.  This
agreement improves as $r$ increases.  In Figs.  \ref{betaprime_Nadj1_figure}
and \ref{betaprime_Nadj2_figure} we present plots of the expansions of
$\beta'_{IR}$ in powers of $\Delta_r$ and in powers of $\Delta_{Adj}$ for these
$FR'$ theories with $R'=Adj$ and $N_{Adj}=1, \ 2$. 

As before for the anomalous dimensions of fermion bilinears, it is of interest
to compare these results in the LNN limit with the results from
Ref. \cite{dexm} for specific values of $N_c$ and $N_F$.  Again, we pick
$N_c=3$ and $N_F=10$, for which the appropriate comparison is with the LNN
values with $r=10/3$.  We can compare these with the values that we obtain in
the LNN limit for the case $N_{Adj}=1$ (for $N_{Adj}=2$, this value of $r$
exceeds $r_u=3/2$).  The values in the six columns of Table
\ref{betaprime_lnn_values_Nadj_eq_1} for $r=10/3$ are $1.70 \times 10^{-3}$,
$1.68 \times 10^{-3}$, $1.79 \times 10^{-3}$, $1.79 \times 10^{-3}$, $1.79
\times 10^{-3}$, and $1.79 \times 10^{-3}$, to be compared with the values in
the corresponding six columns of Table IX of Ref. \cite{dexm}, namely $1.75
\times 10^{-3}$, $1.73 \times 10^{-3}$, $1.84 \times 10^{-3}$, $1.83 \times
10^{-3}$, $1.84 \times 10^{-3}$, and $1.84 \times 10^{-3}$.  One sees that for
each entry in the respective columns of Table
\ref{betaprime_lnn_values_Nadj_eq_1} and the corresponding Table IX in
Ref. \cite{dexm} the results are similar.  As before, this shows the usefulness
of the calculations in the LNN limit, since they approximately reproduce values
of $\beta'$ to a given order of truncation in the scheme-independent series
expansions in an SU($N_c$) theory with $N_F$ fermions in the fundamental
representation with $N_F/N_c$ equal to $r$.  As was the case for the
$\hat\kappa^{(F)}_j$ and $\bar\kappa^{(f')}_j$, for large but finite $N_f$ and
$N_c$, the approach to the LNN limit is rapid for the $\hat d_j$ and $\bar
d_j$, since the subdominant terms again vanish like $1/N_c^2$.

% ========================================================================

\section{$AT$ Theory} 
\label{at_section} 

In this section we analyze the large-$N_c$ limit of the $AT$ theory, i.e.,
a theory in which both the $f$ and $f'$ fermions are in two-index
representations of SU($N_c$).  For finite $N_c$, there are two types of $AT$
theories, namely one with $R_f \equiv R = Adj$ and $R_{f'} \equiv R'=S_2$ and
one with $R_f \equiv R = Adj$ and $R_{f'} \equiv R'=A_2$.  Since the $S_2$ and
$A_2$ representations have the same large-$N_c$ behavior, the $N_c \to \infty$
limits of both of these theories are the same, with $(R,R')=(Adj,T_2)$, where,
as above, $T_2$ stands for either $S_2$ or $A_2$. This is the reason for our
designation of these as the $AT$ theory. The fermions in the adjoint and $T_2$
representations are denoted $\psi$ and $\chi$. 

% ------------------------------------------------------------------------

\subsection{Relevant Interval of $N_{Adj}$ and $N_{T_2}$ for $AT$ Theory} 
\label{at_interval}

In the $N_c \to \infty$ limit of the $AT$ theory, the asymptotic freedom
condition (\ref{af}) reads
\beq
2N_{Adj} + N_{T_2} < \frac{11}{2} \ . 
\label{af_at2}
\eeq
Hence, for a given value of $N_{Adj}$, $N_{T_2}$ must be less than the upper
bound $N_{T_2,u} = (11/2)-2N_{Adj}$, and for a given value of $N_{T_2}$, 
$N_{Adj}$ must be less than the upper bound $N_{Adj,u}=(11/4)-N_{T_2}/2$. 
Let us envision the theories as being specified by a point in the first
quadrant, with the horizontal axis being $N_{Adj}$ and the vertical axis being
$N_{T_2}$.  The upper boundary of the conformal regime is defined by the line
segment $N_{Adj} + (N_{T_2}/2)=11/4$.  This line segment has slope
\beq
\frac{dN_{T_2}}{dN_{Adj}} = -2 \ . 
\label{at2_upper_line_slope}
\eeq
The expansion variables for the scheme-independent series expansions in the 
$AT$ theory are
\beqs
\check \Delta_{Adj} &=& N_{Adj,u}-N_{Adj} \cr\cr
                    &=& \frac{11-2(2N_{Adj}+N_{T_2})}{4}
\label{check_delta_adj_at}
\eeqs
and
\beqs
\check\Delta_{T_2} &=& N_{T_2,u}-N_{T_2} \cr\cr
                   &=& \frac{11-2(2N_{Adj}+N_{T_2})}{2} \ ,
\label{check_delta_t2_at}
\eeqs
where the $\check\Delta$ notation signifies that we have taken the $N_c \to
\infty$ limit.  Thus, 
\beq
\check\Delta_{T_2} = 2\check\Delta_{Adj} \ . 
\label{check_delta_relation_at}
\eeq
As is evident from Eqs. (\ref{check_delta_adj_at}) and 
(\ref{check_delta_t2_at}), $\check\Delta_{T_2}$ and $\check\Delta_{Adj}$ 
depend on $N_{Adj}$ and $N_{T_2}$ only through the combination 
$2N_{Adj}+N_{T_2}$. 

The condition that the two-loop beta function should have an IR zero is
\beq
2N_{Adj} + N_{T_2} > \frac{17}{8} \ . 
\label{at2_lower_line}
\eeq
The lower boundary of the region where the this condition 
(\ref{at2_lower_line}) is satisfied is the line segment 
$2N_{Adj}+N_{T_2}=17/8$ in the first quadrant.  This line segment has the
same slope of $-2$ as the upper boundary.  The region $I_{IRZ}$ is thus given
by 
\beq
I_{IRZ}: \quad \frac{17}{8} < 2N_{Adj}+N_{T_2} < \frac{11}{2} \ . 
\label{irz_at2}
\eeq
For $N_{Adj}$ and $N_{T_2}$ in the IRZ region, the two-loop ($2\ell$) 
rescaled beta function $\beta_{\xi,2\ell}$ has an IR zero at
\beq
\xi_{IR,2\ell} = \frac{2\pi[11-2(2N_{Adj}+N_{T_2})]}
                {8(2N_{Adj}+N_{T_2})-17} \ . 
\label{xir_2loop_at}
\eeq
Note that the upper and lower boundaries of the IRZ regime, the values
of $\check\Delta_{T_2}$ and $\check\Delta_{Adj}$, and the value of 
$\xi_{IR,2\ell}$ depend on $N_{Adj}$ and $N_{T_2}$ only via the combination
$2N_{Adj}+N_{T_2}$. We will assume that $N_{Adj}$ and $N_{T_2}$ are such that
the theory has an IR zero in the conformal regime.

% -------------------------------------------------------------------------

\subsection{$\gamma_{Adj}$ and $\gamma_{T_2}$ in the $AT$ Theory} 
\label{gamma_at_section}

In the $AT$ theory, the coefficients of both types of fermions have finite
large-$N_c$ limits, We denote $\kappa_j^{(f)} \equiv \kappa^{(Adj)}$ and
$\kappa_j^{(f')} \equiv \kappa^{(T_2)}$. With $R_2$ standing for any of the
three two-index representations $Adj$, $S_2$, and $A_2$, we define
\beq
\check\kappa^{(R_2)}_j = 
\lim_{N_c \to \infty} \kappa^{(R_2)}_j \quad {\rm for} \ R_2 \ , 
\label{check_kappa}
\eeq
so that
\beq
\gamma_{\bar\psi\psi,IR} = \sum_{j=1}^\infty \check\kappa^{(Adj)}_j \, 
\check\Delta_{Adj}^j
\label{gamma_adj_ir_Deltaseries}
\eeq
and 
\beq
\gamma_{\bar\chi\chi,IR} = \sum_{j=1}^\infty \check\kappa^{(T_2)}_j \, 
\check\Delta_{T_2}^j \ . 
\label{gamma_t2_ir_Delta2series}
\eeq
We find that for the $\kappa_j$ coefficients that we have calculated, 
\beqs
\check\kappa^{(T_2)}_j &=& \bigg ( \frac{\lambda_{T_2}}{\lambda_{Adj}} 
\bigg )^j \check\kappa^{(Adj)}_j \cr\cr
&=& 2^{-j} \, \check\kappa^{(Adj)}_j
\label{kappa_at2_relation}
\eeqs
From \cite{dexm}, we have
\beq
\check\kappa^{(Adj)}_1 = 2 \check\kappa^{(T_2)}_1 = \frac{2^2}{3^2} = 0.444444
\label{kappa_1_at_lnc}
\eeq
\beq
\check\kappa^{(Adj)}_2 = 2^2 \check\kappa^{(T_2)}_2 = 
\frac{341}{2 \cdot 3^6} = 0.233882 
\label{kappa_2_at_lnc}
\eeq
\beq
\check\kappa^{(Adj)}_3 = 2^3 \check\kappa^{(T_2)}_3 = 
\frac{61873}{2^3 \cdot 3^{10}} = 0.130978 \ . 
\label{kappa_3_at_lnc}
\eeq
The large-$N_c$ limit for these coefficients in a theory with a single fermion
representation $R=Adj$ was previously considered in Ref. \cite{dex}, and the
$\check\kappa^{(Adj)}_j$, $j=1, \ 2, \ 3$ agree with Eqs. (6.18)-(6.21) in that
paper.

Combining the relation $\check\Delta_{T_2} = 2\check\Delta_{Adj}$ from Eq.
(\ref{check_delta_relation_at}) with the relation $\check\kappa^{(T_2)}_j =
2^{-j} \, \check\kappa^{(Adj)}_j$ from Eq.  (\ref{kappa_at2_relation}), we
derive an interesting symmetry property, namely that, for all the orders
$p=1, \ 2, \ 3$ that we have calculated,
\beq
\gamma_{\bar\psi\psi,IR,\check\Delta_{Adj}^p} = 
\gamma_{\bar\chi\chi,IR,\check\Delta_{T_2}^p} \ . 
\label{gamma_equality_at}
\eeq
That is, for the $\psi$ field in the $Adj$ representation and the $\chi$ field
in either the $S_2$ or $A_2$ representation, the $N_c \to \infty$ limits of the
scheme-independent series expansions for the anomalous dimensions of the
corresponding bilinear operators, $\gamma_{\bar\psi\psi,IR}$ and
$\gamma_{\bar\chi\chi,IR}$, are equal to each other at each order that we have
calculated. Furthermore, since the only dependence on $N_{Adj}$ and $N_{T_2}$
enters via the combination $2N_{Adj}+N_{T_2}$, the anomalous dimensions in
Eq. (\ref{gamma_equality_at}) also depend on $N_{Adj}$ and $N_{T_2}$ only
through the combination $2N_{Adj}+N_{T_2}$.  In Table \ref{gamma_at_table} we
list values of $\gamma_{\bar\psi\psi,IR,\check\Delta_{Adj}^p} =
\gamma_{\bar\chi\chi,IR,\check\Delta_{T_2}^p}$ for $p=1, \ 2, \ 3$ in the $AT$
theory for some illustrative values of $N_{Adj}$ and $N_{T_2}$.  As an example
of the dependence on $2N_{Adj}+N_{T_2}$, the values of
$\gamma_{\bar\psi\psi,IR,\check\Delta_{Adj}^p}$ for the theories with
$(N_{Adj},N_{T_2})=(1,3)$ and $(N_{Adj},N_{T_2})=(2,1)$ are the same.

It is of interest to consider the correction terms to the $N_c \to \infty$
limit in this theory.  The coefficients $\kappa^{(Adj)}_j$ with $j=1,2$ are
independent of $N_c$ and hence are equal to their $N_c \to \infty$ limits
$\check\kappa^{(Adj)}_j$ with $j=1,2$.  For $\kappa^{(Adj)}_3$, in a theory
with fermions in only a single representation, $R=Adj$, we recall that (see
Eq. (6.20) in \cite{dex})
\beq
\kappa^{(Adj)}_3 = \frac{61873 - 42624N_c^{-2}}{2^3 \cdot 3^{10}} \quad 
(\rm one \ fermion \ rep.) \ , 
\label{kappa3_adj}
\eeq
so the correction term to the $N_c \to \infty$ limit is proportional to
$1/N_c^2$.  In contrast, we find that the corrections to the $N_c \to \infty$
limits (\ref{kappa_1_at_lnc})-(\ref{kappa_3_at_lnc}) in the $AT$ theory involve
terms proportional to $1/N_c$ rather than $1/N_c^2$.  Consequently, the
approach to the $N_c = \infty$ limit in the $AT$ theory is slower than the
approach to the LNN limit in the $FR'$ theory, since in the latter case the
correction terms are proportional to $1/N_c^2$.

% -------------------------------------------------------------------------

\subsection{$\beta'_{IR}$ Series Expansions in the $AT$ Theory}
\label{betaprime_at_section} 

In the $N_c \to \infty$ limit of the $AT$ theory, the coefficients $d_j$ and 
$\tilde d_j$ in the scheme-independent series expansions for $\beta'_{IR}$ 
are finite.  In accord with our labelling convention that $R_f = Adj$ and
$R_{f'}=T_2$, we denote 
$d_j \equiv d^{(Adj)}_j$ and $\tilde d_j \equiv d^{(T_2)}_j$, and define 
\beq
\check d^{(Adj)}_j = \lim_{N_c \to \infty} d^{(Adj)}_j
\label{check_dj_adj}
\eeq
and
\beq
\check d^{(T_2)}_j = \lim_{N_c \to \infty} d^{(T_2)}_j \ , 
\label{check_dj_t2}
\eeq
so that in this $N_c \to \infty$ limit, the two equivalent scheme-independent 
expansions for $\beta'_{IR}$ are
\beq
\beta'_{IR} = \sum_{j=2}^\infty \check d^{(Adj)}_j \check\Delta_{Adj}^j 
\label{betaprime_series_at}
\eeq
and
\beq
\beta'_{IR} = \sum_{j=2}^\infty \check d^{(T_2)}_j \check\Delta_{T_2}^j \ . 
\label{betaprime_series_at2}
\eeq

For the cases $j=2, \ 3, \ 4$ that we have calculated, we find 
\beqs
\check d^{(T_2)}_j &=& \bigg ( \frac{\lambda_{T_2}}{\lambda_{Adj}} \bigg )^j 
\check d^{(Adj)}_j \cr\cr 
&=& 2^{-j} \, \check d^{(Adj)}_j \ . 
\label{check_dj_rel}
\eeqs
We calculate 
\beq
\check d^{(Adj)}_2 = 2^2 \check d^{(T_2)}_2 = \frac{2^4}{3^4} = 0.197531 \ , 
\label{check_d2_adj}
\eeq
\beq
\check d^{(Adj)}_3 = 2^3 \check d^{(T_2)}_3 = \frac{2^8}{3^7} = 0.117055 \ , 
\label{check_d3_adj}
\eeq
\beq
\check d^{(Adj)}_4 = 2^4 \check d^{(T_2)}_4 = \frac{46871}{2^2 \cdot 3^{12}} = 
0.0220490 \ . 
\label{check_d4_adj}
\eeq

Again, combining the relation 
$\check\Delta_{T_2} = 2\check\Delta_{Adj}$ from Eq.
(\ref{check_delta_relation_at}) with the relation $\check d^{(T_2)}_j =
2^{-j} \, \check d^{(Adj)}_j$ from Eq. (\ref{check_dj_rel}), we find a second
symmetry property characterizing the $N_c \to \infty$ limit of the $AT$ theory,
namely that, for all the orders $p=1, \ 2, \ 3$ that we have calculated,
\beq
\beta'_{IR,\check\Delta_{Adj}^p} = 
\beta'_{IR,\check\Delta_{T_2}^p} \ . 
\label{betaprime_equality_at}
\eeq
We thus write these as $\beta'_{IR,\check\Delta_{R_2}^p}$, where $R_2$ stands
for either $Adj$ or $T_2$.  As discussed in \cite{dexm}, these two
scheme-independent expansions for $\beta'_{IR}$ are equivalent, and here they
are actually identically equal to each order that we have calculated.  As was
the case with the anomalous dimensions of the fermion bilinears, since the only
dependence on $N_{Adj}$ and $N_{T_2}$ enters via the combination
$2N_{Adj}+N_{T_2}$, the scheme-independent series expansion for $\beta'$
depends on $N_{Adj}$ and $N_{T_2}$ only through the combination
$2N_{Adj}+N_{T_2}$.  In Table \ref{betaprime_at_table} we list values of
$\beta'_{IR,\check\Delta_{R_2}^p}$ for $p=2, \ 3, \ 4$ in the $AT$ theory for
some illustrative values of $N_{Adj}$ and $N_{T_2}$.  As another example of the
dependence on $2N_{Adj}+N_{T_2}$, the values of
$\beta'_{IR,\check\Delta_{Adj}^p}$ for the theories with
$(N_{Adj},N_{T_2})=(1,3)$ and $(N_{Adj},N_{T_2})=(2,1)$ are the same.  As with
the $\kappa^{(Adj)}_j$ and the $\kappa^{(T_2)}_j$ coefficients, we find that
the leading-order corrections to the $N_c \to \infty$ limit are proportional to
$1/N_c$. In Figs. 

% ========================================================================

\section{Conclusions}
\label{conclusion_section} 

In this paper we have calculated limiting forms of scheme-independent series
expansions for the anomalous dimensions of gauge-invariant bilinear fermion
operators and of $\beta'$ evaluated at an infrared fixed point of the
renormalization group in asymptotically free SU($N_c)$ gauge theories.  We have
first studied a theory denoted $FR'$ with $N_F$ fermions in the fundamental
representation and $N_{f'}$ fermions in the adjoint, or symmetric or
antisymmetric rank-2 tensor representations, in the limit in which $N_c \to
\infty$ and $N_F \to \infty$ with the ratio $r=N_F/N_c$ fixed and
finite. Secondly, we have studied the $N_c \to \infty$ limit of a theory with
fermions in the adjoint and symmetric or antisymmetric rank-2 tensor
representations, denoted the $AT$ theory. We have shown how these limits yield
useful simplifications of the general results in \cite{dexm}. We have also
determined the nature of the approaches to the respective LNN and $N_c \to
\infty$ limits in the $FR'$ and $AT$ theories. Our results further elucidate
the interesting and fundamental question of the properties of a conformal field
theory, s pecifically, an asymptotically free gauge theory at a conformal
infrared fixed point of the renormalization group

% =======================================================================

\begin{acknowledgments}

This research was supported in part by the Danish National
Research Foundation grant DNRF90 to CP$^3$-Origins at SDU (T.A.R.) and 
by the U.S. NSF Grant NSF-PHY-16-1620628 (R.S.) 

\end{acknowledgments}

% =======================================================================

% ========================================================================

\newpage

% ==============================================================

% table 1 
\begin{table}
  \caption{\footnotesize{Values of $r_\ell$ and $r_u$ as functions of 
  $N_{f'}$ for $R'=Adj$ and $R'=T_2$ ($S_2$ or $A_2$) in the LNN limit of the
  $FR'$ theory.  As noted in the text, since the 
  adjoint representation is self-conjugate, half-integral values of 
  $N_{Adj}$ are allowed, corresponding to $2N_{Adj}$ Majorana fermions.}}
\begin{center}
\begin{tabular}{|c|c|c|} \hline\hline
$R'$  & $r_\ell$ & $r_u$ \\ 
\hline\hline
$N_{Adj}=1/2$  & 1.385   & 4.50 \\
$N_{Adj}=1  $  & 0.154   & 3.50 \\
$N_{Adj}=3/2$  & 0       & 2.50 \\
$N_{Adj}=2$    & 0       & 1.50 \\
\hline
$N_{T_2}=1$    & 1.385   & 4.50 \\
$N_{T_2}=2$    & 0.154   & 3.50 \\
$N_{T_2}=3$    & 0       & 2.50 \\
$N_{T_2}=4$    & 0       & 1.50 \\
$N_{T_2}=5$    & 0       & 0.50 \\
\hline\hline
\end{tabular}
\end{center}
\label{rinterval}
\end{table}

% ========================================================================

% table 2 
\begin{table}
  \caption{\footnotesize{Values of $\hat\kappa^{(F)}_j$ with 
    $j=1, \ 2, \ 3$ in the LNN limit of the $FR'$ theory with 
    $R'=Adj$, as a function of $N_{Adj}$. 
   (As noted in the text, since the adjoint representation is self-conjugate, 
   half-integral values of $N_{Adj}$ are allowed, corresponding to $2N_{Adj}$ 
   Majorana fermions.)  The notation $a$e-n means $10^{-n}$.
    See Table \ref{rinterval} for relevant ranges of
   $N_{Adj}$ as a function of $r$.}}
\begin{center}
\begin{tabular}{|c|c|c|c|} \hline\hline
$N_{Adj}$ & $\hat\kappa^{(F)}_1$ 
          & $\hat\kappa^{(F)}_2$ 
          & $\hat\kappa^{(F)}_3$ \\ 
\hline\hline
1/2       & 0.148  & 0.0339  & 0.718e-2  \\
1         & 0.138  & 0.0307  & 0.642e-2  \\
3/2       & 0.129  & 0.0278  & 0.546e-2  \\
2         & 0.121  & 0.0253  & 0.480e-2  \\
\hline\hline
\end{tabular}
\end{center}
\label{kappahat_values}
\end{table}

% =======================================================================

% table 3 
\begin{table}
  \caption{\footnotesize{Values of $\bar\kappa^{(Adj)}_j$ 
  with $j=1, \ 2, \ 3$ in the LNN limit of the $FR'$ theory with $R'=Adj$ and 
  $N_{Adj}=1$, as a function of $r$.  See Table \ref{rinterval} 
  for relevant ranges of $r$ as a function of $N_{Adj}$.}}
\begin{center}
\begin{tabular}{|c|c|c|c|} \hline\hline
$r$ & $\bar\kappa^{(Adj)}_1$
    & $\bar\kappa^{(Adj)}_2$
    & $\bar\kappa^{(Adj)}_3$ \\ 
\hline\hline
0.2   & 0.449   & 0.238   & 0.1345  \\
0.4   & 0.4545  & 0.243   & 0.138   \\
0.6   & 0.460   & 0.248   & 0.142   \\
0.8   & 0.465   & 0.253   & 0.146   \\
1.0   & 0.471   & 0.2575  & 0.150   \\
1.2   & 0.476   & 0.263   & 0.154   \\
1.4   & 0.482   & 0.268   & 0.159   \\
1.6   & 0.488   & 0.273   & 0.163   \\
1.8   & 0.494   & 0.279   & 0.168   \\
2.0   & 0.500   & 0.285   & 0.173   \\
2.2   & 0.506   & 0.291   & 0.178   \\
2.4   & 0.513   & 0.297   & 0.178   \\
2.6   & 0.519   & 0.303   & 0.189   \\
2.8   & 0.526   & 0.310   & 0.194   \\
3.0   & 0.533   & 0.316   & 0.200   \\
3.2   & 0.541   & 0.323   & 0.206   \\
3.4   & 0.548   & 0.330   & 0.213   \\
\hline\hline
\end{tabular}
\end{center}
\label{kappabar_values}
\end{table}

% =======================================================================

% table 4
\begin{table}
\caption{\footnotesize{Values of the anomalous dimension
 $\gamma_{\bar\psi\psi,IR,\Delta_r^p}$,
 calculated to order $p=1, \ 2, \ 3$ and evaluated at the IR fixed point
in the LNN limit of the $FR'$ theory with $R'=Adj$ and $N_{Adj}=1$, 
as a function of $r$. Here, $\Delta_r=(7-2r)/2$ 
and $\psi$ is the fermion in the $F$ representation. See Table \ref{rinterval}
for relevant range of $r$. See text for further discussion.}}
\begin{center}
\begin{tabular}{|c|c|c|c|} \hline\hline
$r$   & $\gamma_{\bar\psi\psi,IR,\Delta_r}$ &
        $\gamma_{\bar\psi\psi,IR,\Delta_r^2}$ &
        $\gamma_{\bar\psi\psi,IR,\Delta_r^3}$ \\
\hline\hline
0.2   & 0.455   & 0.789   & 1.014   \\
0.4   & 0.428   & 0.722   & 0.908   \\
0.6   & 0.400   & 0.658   & 0.810   \\
0.8   & 0.372   & 0.596   & 0.719   \\
1.0   & 0.345   & 0.5365  & 0.634   \\
1.2   & 0.317   & 0.479   & 0.555   \\
1.4   & 0.290   & 0.425   & 0.483   \\
1.6   & 0.262   & 0.373   & 0.416   \\
1.8   & 0.234   & 0.323   & 0.354   \\
2.0   & 0.207   & 0.276   & 0.297   \\
2.2   & 0.179   & 0.231   & 0.245   \\
2.4   & 0.152   & 0.189   & 0.197   \\
2.6   & 0.124   & 0.149   & 0.154   \\
2.8   & 0.0966  & 0.112   & 0.114   \\
3.0   & 0.0690  & 0.0766  & 0.0774  \\
3.2   & 0.0414  & 0.0441  & 0.0443  \\
3.333 & 0.0230  & 0.0238  & 0.0239  \\
3.4   & 0.01379 & 0.01410 & 0.01411  \\
\hline\hline
\end{tabular}
\end{center}
\label{gamma_fund_lnn_Nf2_eq_1}
\end{table}

% =====================================================================

% table 5
\begin{table}
  \caption{\footnotesize{Values of the anomalous dimension
      $\gamma_{\bar\chi\chi,IR,\Delta_F^p}$,
      calculated to order $p=1, \ 2, \ 3$ and evaluated at the IR fixed point
      in the LNN limit of the $FR'$ theory with $R'=Adj$ and $N_{Adj}=1$,
       as a function of  value of $r$. 
      Here, $\check\Delta_{Adj}=(7-2r)/4$, and $\chi$ is the fermion 
      in the $Adj$ representation. See text for further discussion.}}
\begin{center}
\begin{tabular}{|c|c|c|c|} \hline\hline
$r$   & $\gamma_{\bar\chi\chi,IR,\check\Delta_{Adj}}$ &
        $\gamma_{\bar\chi\chi,IR,\check\Delta_{Adj}^2}$ &
        $\gamma_{\bar\chi\chi,IR,\check\Delta_{Adj}^3}$ \\ 
\hline\hline
0.2   & 0.742  & 1.390  & 1.995   \\
0.4   & 0.705  & 1.288  & 1.803   \\
0.6   & 0.667  & 1.187  & 1.620   \\
0.8   & 0.628  & 1.088  & 1.447   \\
1.0   & 0.588  & 0.991  & 1.284   \\
1.2   & 0.548  & 0.895  & 1.130   \\
1.4   & 0.506  & 0.801  & 0.985   \\
1.6   & 0.463  & 0.710  & 0.850   \\
1.8   & 0.420  & 0.621  & 0.724   \\
2.0   & 0.375  & 0.535  & 0.608   \\
2.2   & 0.329  & 0.452  & 0.501   \\
2.4   & 0.282  & 0.372  & 0.402   \\
2.6   & 0.234  & 0.295  & 0.312   \\
2.8   & 0.184  & 0.222  & 0.230   \\
3.0   & 0.133  & 0.153  & 0.156   \\
3.2   & 0.0811 & 0.0884 & 0.0891  \\
3.333 & 0.04545& 0.0477 & 0.04785 \\
3.4   & 0.02740 & 0.02822 & 0.02825  \\
\hline\hline
\end{tabular}
\end{center}
\label{gamma_adj_lnn_Nf2_eq_1}
\end{table}

% ======================================================================

% table 6
\begin{table}
\caption{\footnotesize{Values of the anomalous dimension
 $\gamma_{\bar\psi\psi,IR,\Delta_r^p}$,
 calculated to order $p=1, \ 2, \ 3$ and evaluated at the IR fixed point
in the LNN limit of the $FR'$ theory with $R'=Adj$ and $N_{Adj}=2$, 
as a function of $r$. Here, $\Delta_r=(3-2r)/2$ 
and $\psi$ is the fermion in the $F$ representation. See Table \ref{rinterval}
for relevant range of $r$. See text for further discussion.}}
\begin{center}
\begin{tabular}{|c|c|c|c|} \hline\hline
$r$   & $\gamma_{\bar\psi\psi,IR,\Delta_r}$ &
        $\gamma_{\bar\psi\psi,IR,\Delta_r^2}$ &
        $\gamma_{\bar\psi\psi,IR,\Delta_r^3}$ \\
\hline\hline
0.2   & 0.158   & 0.200   & 0.211   \\
0.4   & 0.133   & 0.164   & 0.170   \\
0.6   & 0.109   & 0.130   & 0.133   \\
0.8   & 0.0848  & 0.0972  & 0.0989  \\
1.0   & 0.0606  & 0.0669  & 0.0675  \\
1.2   & 0.0364  & 0.0386  & 0.0388  \\
1.4   & 0.0121  & 0.0124  & 0.0124   \\
\hline\hline
\end{tabular}
\end{center}
\label{gamma_fund_lnn_Nf2_eq_2}
\end{table}

% =====================================================================

% table 7
\begin{table}
\caption{\footnotesize{Values of the anomalous dimension
 $\gamma_{\bar\chi\chi,IR,\check\Delta_{Adj}^p}$,
 calculated to order $p=1, \ 2, \ 3$ and evaluated at the IR fixed point
in the LNN limit of the $FR'$ theory with $R'=Adj$ and $N_{Adj}=2$, 
as a function of  value of $r$. 
Here, $\check\Delta_{Adj}=(3-2r)/4$, and $\chi$ is the fermion 
in the $Adj$ representation. See text for further discussion.}}
\begin{center}
\begin{tabular}{|c|c|c|c|} \hline\hline
$r$   & $\gamma_{\bar\chi\chi,IR,\check\Delta_{Adj}}$ &
                    $\gamma_{\bar\chi\chi,IR,\check\Delta_{Adj}^2}$ &
                    $\gamma_{\bar\chi\chi,IR,\check\Delta_{Adj}^3}$ \\ 
\hline\hline
0.2   & 0.292  & 0.393  & 0.430   \\
0.4   & 0.250  & 0.323  & 0.430   \\
0.6   & 0.207  & 0.257  & 0.270   \\
0.8   & 0.163  & 0.194  & 0.200   \\
1.0   & 0.118  & 0.134  & 0.136   \\
1.2   & 0.0714 & 0.0773 & 0.0779  \\
1.4   & 0.0241 & 0.0248 & 0.0248  \\
\hline\hline
\end{tabular}
\end{center}
\label{gamma_adj_lnn_Nf2_eq_2}
\end{table}

% =========================================================================

% figure 1 

% ========================================================================

% figure 2

% =================================================================

% figure 3

% =================================================================

% figure 4

% =================================================================

% table 8
\begin{table}
\caption{\footnotesize{Values of $\beta'_{IR}$ as calculated to order
$O(\Delta_r^p)$ via Eq. (\ref{betaprime_ir_Deltaseries_lnn}),
denoted $\beta'_{IR,\Delta_r^p}$ and to order 
$O(\check\Delta_{Adj}^p)$ via Eq. (\ref{betaprime_ir_Deltaseries2_lnn}),
denoted $\beta'_{IR,\Delta_{Adj}^p}$, with $p=2, \ 3, \ 4$, in the LNN limit 
of the $FR'$ theory with
$R'=Adj$ and $N_{Adj}=1$, as functions of $r$. Here 
$\Delta_r = 2\check\Delta_{Adj} = (7-2r)/2$. 
The notation $a$e-n means $a \times 10^{-n}$.}}
\begin{center}
\begin{tabular}{|c|c|c|c|c|c|c|} \hline\hline
$r$    & $\beta'_{IR,\Delta_r^2}$ & 
         $\beta'_{IR,\check\Delta_{Adj}^2}$ &
         $\beta'_{IR,\Delta_r^3}$ & 
         $\beta'_{IR,\check\Delta_{Adj}^3}$ &
         $\beta'_{IR,\Delta_r^4}$ & 
         $\beta'_{IR,\check\Delta_{Adj}^4}$ \\
\hline\hline
0.2   & 0.668    & 0.544  & 1.326   & 1.0815 & 1.192   & 1.2475   \\
0.4   & 0.589    & 0.485  & 1.135   & 0.941  & 1.031   & 1.0725   \\
0.6   & 0.516    & 0.430  & 0.962   & 0.8115 & 0.883   & 0.9136   \\
0.8   & 0.447    & 0.377  & 0.807   & 0.692  & 0.748   & 0.770    \\
1.0   & 0.383    & 0.327  & 0.669   & 0.583  & 0.625   & 0.641    \\
1.2   & 0.324    & 0.280  & 0.547   & 0.484  & 0.516   & 0.526    \\ 
1.4   & 0.270    & 0.236  & 0.440   & 0.395  & 0.418   & 0.425    \\ 
1.6   & 0.221    & 0.196  & 0.347   & 0.317  & 0.332   & 0.337    \\
1.8   & 0.177    & 0.159  & 0.267   & 0.247  & 0.258   & 0.260    \\
2.0   & 0.138    & 0.125  & 0.200   & 0.1875 & 0.194   & 0.1955   \\
2.2   & 0.104    & 0.0951 & 0.144   & 0.137  & 0.141   & 0.141    \\ 
2.4   & 0.0742   & 0.0689 & 0.0986  & 0.0949 & 0.0969  & 0.09275  \\
2.6   & 0.0497   & 0.04675& 0.0630  & 0.0613 & 0.0623  & 0.0624   \\
2.8   & 0.0300   & 0.0287 & 0.0363  & 0.0357 & 0.03605 & 0.0361   \\
3.0   & 0.0153   & 0.0148 & 0.0176  & 0.0174 & 0.0175  & 0.01755   \\
3.2   & 0.552e-2 & 0.5405e-2&0.601e-2& 0.599e-2& 0.600e-2& 0.600e-3   \\
3.333 & 1.70e-3  & 1.68e-3 & 1.79e-3 & 1.79e-3 & 1.79e-3  & 1.79e-3   \\
3.4   & 0.613e-3 & 0.609e-3& 0.631e-3& 0.631e-3& 0.631e-3& 0.631e-3   \\
\hline\hline
\end{tabular}
\end{center}
\label{betaprime_lnn_values_Nadj_eq_1}
\end{table}

% ===================================================================

% table 9 
\begin{table}
\caption{\footnotesize{Values of $\beta'_{IR}$ as calculated to order
$O(\Delta_r^p)$ via Eq. (\ref{betaprime_ir_Deltaseries_lnn}),
denoted $\beta'_{IR,\Delta_r^p}$ and
to order $O(\Delta_{Adj}^p)$ via Eq. (\ref{betaprime_ir_Deltaseries2_lnn}),
denoted $\beta'_{IR,\check\Delta_{Adj}^p}$, with $p=2, \ 3, \ 4$, 
in the LNN limit of the $FR'$ theory with
$R'=Adj$ and $N_{Adj}=2$, as functions of $r$. Here 
$\Delta_r = 2\check\Delta_{Adj}=(3-2r)/2$. 
The notation $a$e-n means $a \times 10^{-n}$.}}
\begin{center}
\begin{tabular}{|c|c|c|c|c|c|c|} \hline\hline
$r$    & $\beta'_{IR,\Delta_r^2}$ & 
         $\beta'_{IR,\check\Delta_{Adj}^2}$ &
         $\beta'_{IR,\Delta_r^3}$ & 
         $\beta'_{IR,\check\Delta_{Adj}^3}$ &
         $\beta'_{IR,\Delta_r^4}$ & 
         $\beta'_{IR,\check\Delta_{Adj}^4}$ \\
\hline\hline
0.2   & 0.0910   & 0.0844   & 0.122    & 0.117    &  0.121    & 0.121  \\
0.4   & 0.0652   & 0.0611   & 0.0840   & 0.0815   & 0.0835    & 0.0836  \\
0.6   & 0.0436   & 0.0414   & 0.05395  & 0.0528   & 0.0537    & 0.0537  \\
0.8   & 0.0264   & 0.0253   & 0.03125  & 0.0308   & 0.0312    & 0.0312  \\
1.0   & 0.0135   & 0.0137   & 0.0152   & 0.151    & 0.0152    & 0.0152   \\
1.2   & 0.485e-2 & 0.476e-2 & 0.523e-2 & 0.522e-2 & 0.523e-2  & 0.523e-2  \\ 
1.4   & 0.539e-3 & 0.535e-3 & 0.553e-3 & 0.553e-3 & 0.0553e-3 & 0.553e-3  \\ 
\hline\hline
\end{tabular}
\end{center}
\label{betaprime_lnn_values_Nadj_eq_2}
\end{table}

% =========================================================================

% figure 5 

% =========================================================================

% figure 6 

% =======================================================================

% table 10 
\begin{table}
\caption{\footnotesize{Values of the anomalous dimension
 $\gamma_{\bar\psi\psi,IR,\check\Delta_{Adj}^p} = 
 \gamma_{\bar\chi\chi,IR,\check\Delta_{T_2}^p}$, 
 calculated to order $p=1, \ 2, \ 3$ and evaluated at the IR fixed point
in the $AT$ theory for illustrative values of $N_{Adj}$ and $N_{T_2}$.
The respective 
entries are identical for the $(N_{Adj},N_{T_2})=(1,3)$ and $(2,1)$, 
and hence the latter are not shown.}}
\begin{center}
\begin{tabular}{|c|c|c|c|c|c|} \hline\hline
$N_{Adj}$ & $N_{T_2}$ & $2N_{Adj}+N_{T_2}$ & 
$\gamma_{\bar\psi\psi,IR,\check\Delta_{Adj}}$ &
$\gamma_{\bar\psi\psi,IR,\check\Delta_{Adj}^2}$ &
$\gamma_{\bar\psi\psi,IR,\check\Delta_{Adj}^3}$ \\ 
\hline\hline
1 & 2 & 4 & 0.333  & 0.465  & 0.520   \\
1 & 3 & 5 & 0.111  & 0.126  & 0.128   \\
\hline\hline
\end{tabular}
\end{center}
\label{gamma_at_table}
\end{table}

% =======================================================================

% table 11 
\begin{table}
  \caption{\footnotesize{Values of $\beta'_{IR}$ as calculated to order
      $O(\check\Delta_{Adj}^p)$, denoted $\beta'_{IR,\check\Delta_{Adj}^p}$ and
      to order $O(\check\Delta_{T_2}^p)$, denoted
      $\beta'_{IR,\check\Delta_{T_2}^p}$, with $p=2, \ 3, \ 4$, in the AT 
      theory for illustrative values of $N_{Adj}$ and 
      $N_{T_2}$. As discussed in the text,
      $\beta'_{IR,\check\Delta_{Adj}^p}=\beta'_{IR,\check\Delta_{T_2}^p}$, 
      so we denote these as
      $\beta'_{IR,\check\Delta_{R_2}^p}$, where $R_2$ stands for either $Adj$ 
      or $T_2$. The respective 
      entries are identical for the $(N_{Adj},N_{T_2})=(1,3)$ and $(2,1)$, 
      and hence the latter are not shown.}}
\begin{center}
\begin{tabular}{|c|c|c|c|c|c|} \hline\hline
$N_{Adj}$ & $N_{T_2}$ & $2N_{Adj}+N_{T_2}$ & 
$\beta'_{IR,\check\Delta_{R_2}^2}$ & 
$\beta'_{IR,\check\Delta_{R_2}^3}$ & 
$\beta'_{IR,\check\Delta_{R_2}^4}$ \\
\hline\hline
1 & 2 & 4 & 0.111  & 0.1605  & 0.1675  \\
1 & 3 & 5 & 0.0123 & 0.0142  & 0.0143  \\
\hline\hline
\end{tabular}
\end{center}
\label{betaprime_at_table}
\end{table}

% ======================================================================


\begin{thebibliography}{99}
\bibitem{fm}
Taking the fermions to be massless does not involve any loss of generality,
because a fermion with a nonzero mass $m_0$ would be
integrated out of the low-energy effective field theory at Euclidean momentum
scales $\mu < m_0$, and hence would not affect the properties of the
theory at the IRFP.
\bibitem{dexnote}
%
  In Eqs. (3.3) and (5.7) of \cite{dex}, $b_\ell^{(r)}$ should be $((-1)^r/r!)
  b_\ell^{(r)}$ and in Eq. (4.5), $c_{{\cal O},\ell}^{(r)}$ should be
  $((-1)^r/r!)c_{{\cal O},\ell}^{(r)}$.  These were just switches in notation 
  in the text and did not affect the calculations or results.
\bibitem{casimir}
%
We recall the general definitions of these group invariants. Denote $T_R$ as a 
generator of the Lie algebra of a group $G$. Then the quadratic Casimir
invariant $C_2(R)$ is defined by $T^a_RT^a_R = C_2(R) I$, 
where $I$ is the $d_R \times d_R$ identity matrix, and the trace invariant
$T(R)$  is defined by ${\rm Tr}_R(T^a_R T^b_R) = T(R)\delta_{ab}$, where 
$1 \le a \le o(G)$, with $o(G)$ the order of the group. 
We write $C_A=C_2(Adj)$, $C_f = C_2(R_f)$, and $C_{f'} = C_2(R_{f'})$. 
For SU($N_c$), $C_2(Adj)=T(Adj)=N_c$; $C_2(T_2)=(N_c \pm 2)(N_c \mp 1)/N_c$; 
and $T(T_2) = (N_c \pm 2)/2$, where the $+$ and $-$ signs apply for $T_2=S_2$ 
and $A_2$, respectively.
\end{thebibliography}
\end{document}